\documentclass[%
aps,
pra,
superscriptaddress,
tightenlines,
showpacs,showkeys,
a4paper,
12pt,
longbibliography,
notitlepage
]{revtex4-1}
\usepackage[english]{babel}
\usepackage{amssymb,amsmath,stmaryrd,array,esint}

\usepackage{graphicx}
\usepackage{subfig}

\makeatletter

\newcommand{\sca}[2]{\ensuremath{\bigl({#1}\cdot{#2}\bigr)}}
\newcommand{\avr}[1]{\ensuremath{\langle{#1}\rangle}}

\newcommand{\cnj}[1]{{#1}^{\ast}}
\newcommand{\hcnj}[1]{{#1}^{\dagger}}

\newcommand{\prt}[1]{\partial_{#1}}

\newcommand{\bnbl}{\boldsymbol{\nabla}}

\newcommand{\diag}{\mathop{\rm diag}\nolimits}

\renewcommand{\Re}{\mathop{\rm Re}\nolimits}
\renewcommand{\Im}{\mathop{\rm Im}\nolimits}

 \newcommand{\bs}[1]{\boldsymbol{#1}}
 \newcommand{\vc}[1]{\mathbf{#1}}
 \newcommand{\mvc}[1]{\mathbf{#1}}
 \newcommand{\uvc}[1]{\hat{\mathbf{#1}}}
 
 \newcommand{\ind}[1]{\mathrm{#1}}

\newcommand{\dd}{\mathrm{d}}

 \newcommand{\e}{\mathrm{e}}
 \newcommand{\ee}{\mathrm{e}}

\makeatother

\begin{document}
\DeclareGraphicsExtensions{.pdf,.png,.eps}
\title{
Optical trapping by Laguerre-Gaussian beams:\\
Symmetries, stability and equilibria 
}

\author{Alexei~D.~Kiselev}
\email[Email address: ]{alexei.d.kiselev@gmail.com}
\affiliation{%
Saint Petersburg National Research University of Information
Technologies, Mechanics and Optics (ITMO University),
Kronverskyi Prospekt 49, 
197101 St Petersburg, Russia
}
\author{Dmytro~O.~Plutenko}
\email[Email address: ]{dmplutenko@gmail.com}
\affiliation{%
 Institute of Physics of National Academy of Sciences of Ukraine,
 prospekt Nauki 46,
 03680 Kiev, Ukraine}
\affiliation{%
 Physical Engineering Teaching Research Center of National Academy of Sciences of Ukraine,
 Kiev, Ukraine}

\date{\today}

\begin{abstract}
By using the method of far-field matching
we obtain the far-field expressions for
the optical (radiation) force
exerted by Laguerre--Gaussian (LG) light beams
on a spherical (Mie) particle
and study
the optical-force-induced dynamics of the scatterer 
near the trapping points represented
by the equilibrium (zero-force) positions.
The regimes of linearized dynamics are
described in terms of the stiffness matrix spectrum
and the damping constant of the ambient medium.
Numerical analysis is performed 
for both non-vortex and optical vortex LG beams.
For the purely azimuthal LG beams,
the dynamics is found to be locally non-conservative 
and is characterized by the presence of
conditionally stable equilibria
(unstable zero-force points that can be stabilized
by the ambient damping).
We also discuss effects related to the Mie resonances
(maxima of the internal field Mie coefficients)
that under certain conditions manifest themselves
as the points changing the trapping properties
of the particles.
\end{abstract}


 \maketitle

\section{Introduction}
\label{sec:intro}

The idea of a mechanical action of light has been 
pursued for hundreds of years.
In the 1970s Ashkin demonstrated the fact that focused
laser beams can be used to trap and control 
dielectric particles,
which also included feedback 
cooling~\cite{Ashkin:prl:1970,Ashkin:bk:2006}. 
Over the past two decades
single-beam optical traps,
that were first developed in~\cite{Ashkin:ol:1986}
and are
commonly known as the optical tweezers,
have become an indispensable tool 
in numerous fields of science 
where optical forces are employed for manipulation, measurements, 
or for creating and controlling new states of matter.

Theoretical approaches to modeling of optical
tweezers 
are typically based on the theory of light scattering~\cite{Nieminen:jqsrt:2014,Salandrino:josab:2012}
and use the methods closely related to
the problem of light scattering by spherically shaped particles 
that dates back to
the more than century-old 
classical exact solution due to Mie~\cite{Mie:1908}.  
A systematic expansion of
the electromagnetic field over vector spherical 
harmonics lies at the heart
of Mie--type theories~\cite{New,Tsang:bk1:2000,Mishchenko:bk:2004,Gouesbet:bk:2011}.  

The
specific form of the expansions known as 
the \textit{T}--matrix ansatz  
has been widely used in the related problem of light scattering
by non-spherical~\cite{Mis:1996,Mish,Mishchenko:bk:2004} 
and optically anisotropic particles~\cite{Kiselev:optsp:2000,Kiselev:pre:2002,
Geng:pre:2004,Novitsky:pra:2008,Qiu:lpr:2010}.
Light scattering from arbitrary shaped
laser beams~\cite{Grehan:aplopt:1986,Gouesbet:josaa:1988,Barton:jap:1988,Barton:jap:2:1989,Schaub:josaa:1992}
has been the key subject of
the Mie--type theory~---~the so-called
{generalized Lorenz--Mie theory} (GLMT)
~---~extended 
to the case of arbitrary incident-beam scattering~\cite{Lock:jqsrt:2009,Gouesbet:bk:2011}.

In GLMT, illuminating beams are described in terms of 
expansions over a set of basis wavefunctions
and the expansion coefficients known as the
\textit{beam shape coefficients}~\cite{Gouesbet:jqsrt:2:2011}.
When the analytical treatment
of laser beams uses approximations such as 
the well-known paraxial approximation~\cite{Lax:pra:1975},
the key difficulty is that 
multipole expansions for 
approximate solutions
of the vector Helmholtz equation (pseudo-fields)
representing the beams
do not exist. 
Therefore, some remodelling procedure must be invoked
to obtain a real radiation field.

Typically, remodelling procedure assume  that 
the actual incident field is equal to the pseudo-field
on a matching surface such as 
a far-field sphere~\cite{Nieminen:jqsrt:2003}, 
the focal plane~\cite{Nieminen:jqsrt:2003,Bareil:josaa:2013},
and a Gaussian reference sphere representing 
a lens~\cite{Hoang:josaa:2012}.
Given the pseudo-field distribution
on the surface,
the beam shape coefficients
then can be evaluated using
either numerical integration
or the one-point matching method~\cite{Nieminen:jqsrt:2003}.

Alternatively,
propagation of a laser beam,
which is known in the paraxial limit,
can be analytically described
without recourse to the paraxial approximation.
In Refs.~\cite{Barnett:optcomm:1994,Duan:josaa:2005,Ness:jmo:2006,Zhou:ol:2006,Zhou:olt:2008}
this strategy has been applied 
to the important case of Laguerre--Gaussian (LG) beams
using different methods.

In recent studies of light scattering by spherical and spheroidal particles
illuminated with LG beams~\cite{Torok:optexp:2007,Jiang:jopt:2012}, 
the analytical results of Ref.~\cite{Ness:jmo:2006}
were used to calculate the beam shape coefficients.
In our previous paper~\cite{Kiselev:pra:2014}
the far-field matching method is
combined with  the results for nonparaxial propagation of
LG beams~\cite{Zhou:ol:2006,Zhou:olt:2008}
to study near-field structures such as nanojets and optical vortices.
Similar method was recently used in Ref.~\cite{Yu:josaa:2015}.

LG beams are important for optical trapping~\cite{Otsu:srep:2014}.
At nonzero azimuthal mode number, they represent 
optical vortex
laser beams exhibiting a helical phase front
and carrying a phase singularity.
The topological charge of the phase singularity
and associated orbital angular momentum
are known to give rise to a number of distinctive phenomena~\cite{Allen:bk:2003,Andrews:bk:2008}.
In particular, 
rotation of trapped spheres by vortex
beams~\cite{Grier:nat:2003,Simpson:josaa:1:2009} 
is a remarkable manifestation of
the non-conservative nature of optical-force-induced dynamics.
The latter implies that,
owing to a scattering contribution to optical force fields, 
the optical forces cannot generally be derived from an underlying
potential.
This has important consequences
for stochastic dynamics of the particles optically trapped by 
LG beams. In particular,
such particles may not approach
thermodynamic equilibrium~\cite{Roichman:prl:2008,Simpson:pre:2010,Ruffner:prl:2012,Siler:optex:2012}.
The fundamental problems of
nonequilibrium statistical physics~\cite{Seifert:rpp:2012,Kiselev:pre:1:2010}
have thus given additional impetus to
the studies on technologically important subject 
related to dynamics of optical trapping.

In this paper we 
deal with the radiation-force-induced dynamics
of Mie scatterers.
Our goal is to examine the regimes of the dynamics
depending on the parameters
characterizing both the LG beam and the scatterer. 
 For this purpose, we
systematically use
the far-field matching method 
supplemented with the symmetry analysis.
The effects related to the non-conservative
character of the dynamics 
and the role of Mie resonances will be of our primary interest. 

The layout of the paper is as follows.  
In Sec.~\ref{sec:t-matrix-approach},
we outline our theoretical approach.
The analytical results
for the beam shape coefficients
of LG beams and the fundamental 
properties
of the far-field angular distributions
are described in
Sec.~\ref{sec:matching}.
The far-field expressions for 
the flux of the Poynting vector
and the optical force
are deduced in Sec.~\ref{subsec:ff-poynting}.
Symmetry analysis for LG beams
is performed in Sec.~\ref{subsec:symmetry}.
The optical-force-induced dynamics
and stability of the zero-force points are
discussed in Sec.~\ref{subsec:dynamics}.
In the remaining part of
Sec.~\ref{sec:results},
we present
the results of numerical computations
such as the stiffness matrix eigenvalues
and the on-axis position of the equilibria
evaluated as a function
of the scatterer size parameter
at different values of the LG beam 
radial and azimuthal
mode numbers.
Technical details on 
separating out
the gradient dependent terms
in 
the expression for the optical
force are relegated to Appendix~\ref{sec:derivation}.
Finally, in Sec.~\ref{sec:conclusions},
we draw our results together and make some concluding
remarks.

\begin{figure*}[!tbh]
\includegraphics[width=50mm]{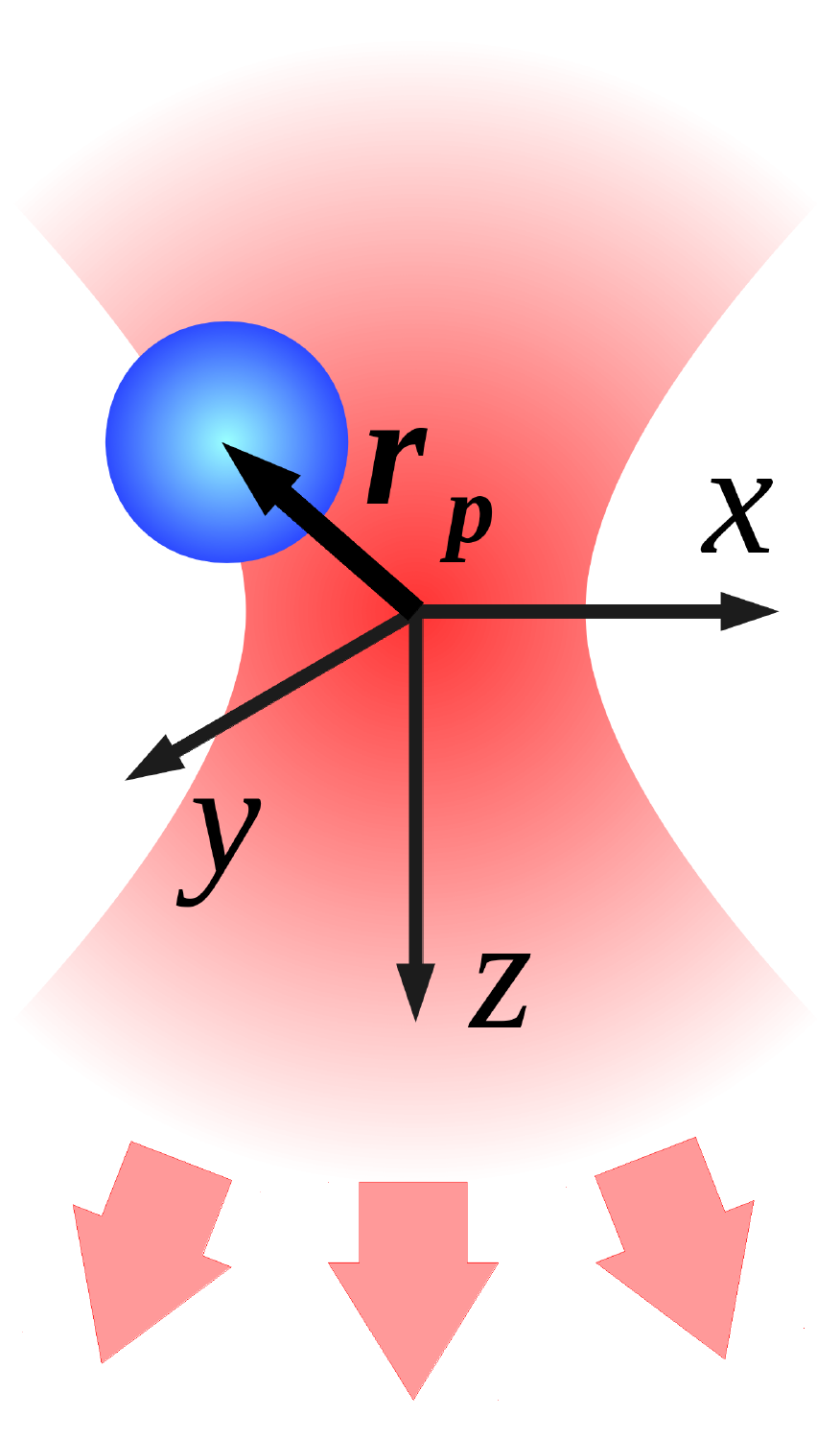}
\caption{%
(Color online)
Mie scatterer (spherical particle) is illuminated with
a focused LG beam propagating along the $z$ axis.
The displacement vector $\vc{r}_p$
determines location of the particle
with respect to the beam waist. 
}
\label{fig:beam}
\end{figure*}

\section{Lorenz--Mie theory: Wave functions and \textit{T}--matrix}
\label{sec:t-matrix-approach}

In this section we 
introduce all necessary notations and
briefly discuss how the properties of 
Mie scattering can be 
described in terms of 
the \textit{T}--matrix~\cite{New,Mishchenko:bk:2004}. 
Our formulation closely follows to the line of our
presentation given in Refs.~\cite{Kiselev:pre:2002,Kiselev:pra:2014}. 

We consider scattering by a spherical particle of radius $R_p$
embedded in a uniform isotropic dielectric medium with dielectric
constant $\epsilon_{\ind{med}}$ and magnetic
permeability $\mu_{\ind{med}}$
(the geometry of light scattering is schematically illustrated in Fig.~\ref{fig:beam}).
The dielectric constant and magnetic
permittivity of the particle are $\epsilon_p$ and $\mu_p$,
respectively.
For a harmonic electromagnetic wave
(time--dependent factor is $\exp\{-i\omega t\}$),
the  Maxwell equations can be written in 
the following form:
\begin{subequations}
  \label{eq:maxwell}
\begin{align}
-i k_i^{-1}\,
\bs{\nabla}\times\vc{E}&=\frac{\mu_i}{n_i} \vc{H}\, ,
\label{eq:maxwell1}\\
i k_i^{-1}\,
\bs{\nabla}\times\vc{H}&= \frac{n_i}{\mu_i} \vc{E},
\quad
i=
\begin{cases}
  \ind{med}, & r>R_p\\
p, & r<R_p
\end{cases}
\label{eq:maxwell2}
\end{align}
\end{subequations}
where 
$n_{\ind{med}}=\sqrt{\epsilon_{\ind{med}}\mu_{\ind{med}}}$
is the refractive index outside the scatterer
(in the ambient medium),
where $r>R_p$ ($i=\ind{med}$)
and 
$k_i=k_{\ind{med}}=n_{\ind{med}}k_{\ind{vac}}$
($k_{\ind{vac}}=\omega/c=2\pi/\lambda$ is the free--space wave number);
$n_{p}=\sqrt{\epsilon_{p}\mu_{p}}$ 
is the refractive index
for the region inside the spherical particle (scatterer), 
where $r<R_p$ ($i=p$) and $k_i=k_p=n_pk_{\ind{vac}}$.
  
The electromagnetic field can 
always be expanded
using the vector spherical harmonic basis~\cite{Biedenharn:bk:1981}.
There are three cases of these expansions that are of particular
interest. They correspond to the incident wave,
$\{\vc{E}_{\ind{inc}},\vc{H}_{\ind{inc}}\}$, the outgoing scattered wave,
$\{\vc{E}_{\ind{sca}},\vc{H}_{\ind{sca}}\}$
and the electromagnetic field inside the scatterer,
$\{\vc{E}_{p},\vc{H}_{p}\}$:
\begin{subequations}
\label{eq:EH}
\begin{align}
& 
\vc{E}_{\alpha}=
\sum_{jm}
\bigl[
\alpha_{jm}^{(\alpha)}\vc{M}_{jm}^{(\alpha)}(\rho_{i},\uvc{r})+
\beta_{jm}^{(\alpha)}\vc{N}_{jm}^{(\alpha)}(\rho_{i},\uvc{r})
\bigr],
\quad
\alpha\in\{\ind{inc}, \ind{sca}, p\}
\label{eq:E_alpha}
\\
& 
\vc{H}_{\alpha}=n_{i}/\mu_{i}\sum_{jm}
\bigl[
\alpha_{jm}^{(\alpha)}\vc{N}_{jm}^{(\alpha)}(\rho_{i},\uvc{r})-
\beta_{jm}^{(\alpha)}\vc{M}_{jm}^{(\alpha)}(\rho_{i},\uvc{r})
\bigr],
\label{eq:H_alpha}
\\
&
  \label{eq:M_jm}
  \vc{M}_{jm}^{(\alpha)}(\rho_i,\uvc{r})=i k_i^{-1}\,\bs{\nabla}\times\vc{N}_{jm}^{(\alpha)}=z_j^{(\alpha)}(\rho_i)\vc{Y}_{jm}^{(m)}(\uvc{r}),
\\
&
  \label{eq:N_jm}
\vc{N}_{jm}^{(\alpha)}(\rho_i,\uvc{r})=-i k_i^{-1}\,\bs{\nabla}\times\vc{M}_{jm}^{(\alpha)}=
\frac{\sqrt{j(j+1)}}{\rho_i}\,
z_j^{(\alpha)}(\rho_i)\, \vc{Y}_{jm}^{(0)}(\uvc{r})+
D z_j^{(\alpha)}(\rho_i) \vc{Y}_{jm}^{(e)}(\uvc{r}),
\\
&
i=
\begin{cases}
  \ind{med}, & \alpha\in\{\ind{inc}, \ind{sca}\}\\
p, & \alpha=p
\end{cases},
\quad
z_j^{(\alpha)}(\rho_i)=
\begin{cases}
  j_j(\rho), & \alpha=\ind{inc}\\
h_j^{(1)}(\rho), & \alpha=\ind{sca}\\
j_j(\rho_p), & \alpha=p\\
\end{cases},
\end{align} 
\end{subequations}
where
$\rho\equiv\rho_{\ind{med}}=k_{\ind{med}}r$,
$\rho_{p}=k_{p} r\equiv n \rho$,
and $n=n_p/n_{\ind{med}}$ is 
the ratio of refractive indexes
also known as the \textit{optical contrast};
$D f(x)\equiv x^{-1}\partial_{x}(x f(x))$
and $\prt{x}$ stands for a derivative with respect to $x$.

According to Ref.~\cite{Kiselev:pre:2002},
the spherical harmonics
can be conveniently expressed in terms of the Wigner 
\textit{D}--functions~\cite{Biedenharn:bk:1981,Varshalovich:bk:1988} 
as follows
\begin{subequations}
\label{eq:Y_D}
\begin{align}
& \vc{Y}_{jm}^{(m)}(\uvc{r})=
N_j /\sqrt{2}\left\{
D_{m,\,-1}^{j\,*}(\uvc{r})\,\vc{e}_{-1}(\uvc{r})-
D_{m,\,1}^{j\,*}(\uvc{r})\,\vc{e}_{+1}(\uvc{r})
\right\}\, ,
  \label{eq:Ym_D}\\
& \vc{Y}_{jm}^{(e)}(\uvc{r})=
N_j /\sqrt{2} \left\{
D_{m,\,-1}^{j\,*}(\uvc{r})\,\vc{e}_{-1}(\uvc{r})+
D_{m,\,1}^{j\,*}(\uvc{r})\,\vc{e}_{+1}(\uvc{r})
\right\}\, ,
  \label{eq:Ye_D}\\
& \vc{Y}_{jm}^{(0)}(\uvc{r})=
N_j
D_{m,\, 0}^{j\,*}(\uvc{r})\,\vc{e}_{0}(\uvc{r})
= Y_{jm}(\uvc{r})\uvc{r},
\quad
N_j=[(2j+1)/4\pi]^{1/2},
  \label{eq:Y0_D}
\end{align}
\end{subequations}
where
$\vc{Y}^{(m)}_{jm}$, 
$\vc{Y}^{(e)}_{jm}$ 
and
$\vc{Y}^{(0)}_{jm}$
are electric,
magnetic and longitudinal harmonics, respectively; 
$\vc{e}_{\pm 1}(\uvc{r})=
\mp (\vc{e}_x(\uvc{r})\pm i \vc{e}_y(\uvc{r}))/\sqrt{2}$;
$\vc{e}_x(\uvc{r})\equiv \vc{e}_{\theta}(\uvc{r}) 
=
(\cos\theta\cos\phi, \cos\theta\sin\phi, -\sin\theta)$,
$\vc{e}_y(\uvc{r})\equiv\vc{e}_{\phi}(\uvc{r})
=(-\sin\phi, \cos\phi, 0)$
are the unit vectors tangential to the sphere;
$\phi$ ($\theta$) is the azimuthal (polar) angle of the unit vector 
$\uvc{r}=\vc{r}/r=
(\sin\theta\cos\phi, \sin\theta\sin\phi, \cos\theta)
\equiv\vc{e}_0(\uvc{r})\equiv\vc{e}_z(\uvc{r})$;
$f(\uvc{r})\equiv f(\phi,\theta)$.
(Hats will denote unit vectors and an asterisk will indicate 
complex conjugation.)

Note that,
for the irreducible representation
of the rotation group with the angular number
$j$, 
the \textit{D}-functions,
$D_{m\nu}^{\,j}(\alpha,\beta,\gamma)=\exp(-i m \alpha)
d_{m\mu}^{\,j}(\beta) \exp(-i \mu \gamma)$, 
give the elements of the rotation matrix
parametrized by the three Euler 
angles~\cite{Biedenharn:bk:1981,Varshalovich:bk:1988}:
$\alpha$, $\beta$ and $\gamma$.
In formulas~\eqref{eq:Y_D} and throughout this paper,
we assume that
$\gamma=0$ and 
$D_{m\nu}^{\,j}(\uvc{r})\equiv D_{m\nu}^{\,j}(\phi,\theta,0)$.
Owing to the
orthogonality relations
for \textit{D}-functions~\cite{Biedenharn:bk:1981,Varshalovich:bk:1988}, 
a set of vector spherical harmonics is 
orthonormal:
\begin{equation}
  \langle \vc{Y}_{jm}^{(\alpha)\,*}(\uvc{r})\cdot
\vc{Y}_{j'm'}^{(\beta)}(\uvc{r})
\rangle_{\uvc{r}}= \delta_{\alpha\beta}\,\delta_{jj'}\,\delta_{mm'}\, .
  \label{eq:Y-orth}
\end{equation}
where
$\displaystyle
\langle\,f\,\rangle_{\uvc{r}}\equiv\int_0^{2\pi}\dd\phi
\int_0^{\pi}\sin\theta\dd\theta\,f(\uvc{r})$.

It can be shown~\cite{Kiselev:pra:2014} that
the vector spherical harmonics~\eqref{eq:Y_D}
can also be recast into the well-known standard form~\cite{Jacks:bk:1999}:
\begin{subequations}
\label{eq:Y_Yjm}
\begin{align}
&
\label{eq:Ym-in-Yjm}
\vc{Y}_{jm}^{(m)}(\uvc{r})=
n_j \vc{L} {Y}_{jm}=-i \uvc{r}\times \vc{Y}_{jm}^{(e)},
\\
&
\label{eq:Ye-in-Yjm}
\vc{Y}_{jm}^{(e)}(\uvc{r})=
n_j r \bs{\nabla} {Y}_{jm} =-i \uvc{r}\times \vc{Y}_{jm}^{(m)},
\quad
n_j\equiv [j(j+1)]^{-1/2},
\end{align}
\end{subequations}
where 
${Y}_{jm}(\uvc{r})\equiv {Y}_{jm}(\phi,\theta)$ is the normalized spherical function;
$\vc{L}=-i \vc{r}\times\bs{\nabla}$ 
is the operator of angular momentum

The vector wave functions, $\vc{M}_{jm}^{(\alpha)}$ and
$\vc{N}_{jm}^{(\alpha)}$,
are the solenoidal solutions of the
vector Helmholtz equation
that
can be derived
(a discussion of the procedure can be found, e.g.,
in Ref.~\cite{Sarkar:pre:1997})
from
the solutions of the scalar
Helmholtz equation
taken in the factorized form:
$\psi_{jm}^{(\alpha)}=n_j z_j^{(\alpha)}(k r) \ind{Y}_{jm}(\uvc{r})$,
where  
$z_j^{(\alpha)}(x)$ is either a spherical Bessel 
function,
$j_j(x)=[\pi/(2x)]^{1/2} J_{j+1/2}(x)$,
or a spherical Hankel function~\cite{Abr}, 
$h_j^{(1,\,2)}(x)=[\pi/(2x)]^{1/2}
H_{j+1/2}^{(1,\,2)}(x)$.

In the far-field region
($\rho\gg 1$), the asymptotic behavior of the spherical Bessel
and Hankel functions is known~\cite{Abr}:
\begin{subequations}
\label{eq:asymp-h-j}
\begin{align}
&
  \label{eq:asymp-hankel1}
i^{j+1}h_j^{(1)}(\rho), i^{j}Dh_j^{(1)}(\rho)
\sim 
\exp(i\rho)/\rho,
\\
&
  \label{eq:asymp-hankel2}
(-i)^{j+1}h_j^{(2)}(\rho), (-i)^{j}Dh_j^{(2)}(\rho)
\sim 
\exp(-i\rho)/\rho,
\
\\
&
  \label{eq:asymp-bessel}
i^{j+1}j_j(\rho),i^{j+1}Dj_{j+1}(\rho)
\sim 
\bigl[\exp(i\rho)-(-1)^j \exp(-i\rho)
\bigr]/(2\rho).
\end{align}
\end{subequations}
So, the spherical Hankel functions of the first kind,
$h_j^{(1)}(\rho)$, describe the outgoing waves,
whereas those of the second kind, $h_j^{(2)}(\rho)$,
represent the incoming waves.
For such waves, 
similar to Eqs.~\eqref{eq:E_alpha}-\eqref{eq:H_alpha}, 
the expansions in vector spherical harmonics
can be written in terms of the vector wave functions:
$\vc{M}_{jm}^{(1,\,2)}$ and $\vc{N}_{jm}^{(1,\,2)}$ 
given in Eqs.~\eqref{eq:M_jm} and~\eqref{eq:N_jm}
with $z_j^{(1,\,2)}=h_j^{(1,\,2)}$.
From the asymptotic relations~\eqref{eq:asymp-hankel1}
and~\eqref{eq:asymp-hankel2},
the vector wave functions
of outgoing and incoming waves
in the far-field region
are given by
\begin{align}
&
  \label{eq:MN_out-asympt}
 \vc{M}_{jm}^{(\ind{out})}
\equiv \vc{M}_{jm}^{(1)}
\sim 
(-i)^{j+1}\frac{\e^{i\rho}}{\rho}
\vc{Y}_{jm}^{(m)},
\quad
 \vc{N}_{jm}^{(\ind{out})}
\equiv \vc{N}_{jm}^{(1)}
\sim 
(-i)^{j}\frac{\e^{i\rho}}{\rho}
\vc{Y}_{jm}^{(e)},
\\
&
  \label{eq:MN_in-asympt}
 \vc{M}_{jm}^{(\ind{in})}
\equiv \vc{M}_{jm}^{(2)}
\sim 
i^{j+1}\frac{\e^{-i\rho}}{\rho}
\vc{Y}_{jm}^{(m)},
\quad
 \vc{N}_{jm}^{(\ind{in})}
\equiv \vc{N}_{jm}^{(2)}
\sim 
i^{j}\frac{\e^{-i\rho}}{\rho}
\vc{Y}_{jm}^{(e)}.
\end{align}

Thus outside the scatterer the optical field is a sum of the
incident wave field with $z_j^{(\ind{inc})}(\rho)=j_j(\rho)=[h_j^{(1)}(\rho)+h_j^{(2)}(\rho)]/2$
and the scattered waves with
$z_j^{(\ind{sca})}(\rho)=h_j^{(1)}(\rho)$ as required by the
Sommerfeld radiation condition.
  The incident field is the field that would exist
without a scatterer and therefore includes
both incoming and outgoing parts
(see Eq.~\eqref{eq:asymp-bessel})
because, when no scattering, 
what comes in must go outwards again.
As opposed to the spherical Hankel functions
that are singular at the origin, 
the incident wave field should be finite everywhere
and thus is described by the regular Bessel functions
$j_j(\rho)$.

Now the incident wave is characterized by amplitudes
$\alpha_{jm}^{(\ind{inc})}$, $\beta_{jm}^{(\ind{inc})}$ and the scattered
outgoing waves are similarly characterized by amplitudes
$\alpha_{jm}^{(\ind{sca})}$, $\beta_{jm}^{(\ind{sca})}$.  
So long as the scattering problem is
linear, the coefficients $\alpha_{jm}^{(\ind{sca})}$ and
$\beta_{jm}^{(\ind{sca})}$ can be written as linear combinations of
$\alpha_{jm}^{(\ind{inc})}$ and $\beta_{jm}^{(\ind{inc})}$:
\begin{gather}
  \alpha_{jm}^{(\ind{sca})}=\sum_{j',m'}\left[\,
T_{jm,\,j'm'}^{\,11}\, \alpha_{j'm'}^{(\ind{inc})}+ 
T_{jm,\,j'm'}^{\,12}\,\beta_{j'm'}^{(\ind{inc})}
\,\right],
\notag
\\
\beta_{jm}^{(\ind{sca})}=\sum_{j',m'}\left[\,
T_{jm,\,j'm'}^{\,21}\, \alpha_{j'm'}^{(\ind{inc})}+ 
T_{jm,\,j'm'}^{\,22}\,\beta_{j'm'}^{(\ind{inc})}
\,\right]\, .
  \label{eq:matr}
\end{gather}
These formulas define the elements of the \textit{\textit{T}--matrix} in the most general case.

In general, the scattering process mixes angular momenta~\cite{Mis:1996}.
The light scattering from uniformly anisotropic 
scatterers~\cite{Kiselev:pre:2002,Kiselev:mclc:2002,Geng:pre:2004,Stout:1:josa:2006}
provides an example of such a scattering process.
In simpler scattering processes, by contrast, such angular momentum
mixing does not take place. 
For example, radial anisotropy keeps intact spherical symmetry of 
the scatterer~\cite{Rot:1973,Kiselev:pre:2002,Qiu:lpr:2010}.
The \textit{T}--matrix of a spherically symmetric scatterer is diagonal
over the angular momenta and the azimuthal numbers:
$T_{jj',mm'}^{nn'}=\delta_{jj'}\delta_{mm'} T_{j}^{nn'}$.

In order to calculate the elements of \textit{T}-matrix
and the coefficients $\alpha_{jm}^{(p)}$ and $\beta_{jm}^{(p)}$, we need to use
continuity of the tangential components of the electric and magnetic
fields as boundary conditions at $r=R_p$ 
($\rho=k_{\ind{med}} R_p\equiv x$).
So, the coefficients of the expansion for the wave field inside the
scatterer, $\alpha_{jm}^{(p)}$ and $\alpha_{jm}^{(p)}$, 
are expressed in terms of the coefficients
describing the incident light as follows
\begin{align}
&
  \label{eq:mie-alp-p}
 \alpha_{jm}^{(p)}\equiv a_j^{(p)}\alpha_{jm}^{(\ind{inc})}=
\frac{-i \alpha_{jm}^{(\ind{inc})}}{%
\mu^{-1} v_j(x) u_j^{\prime}(n x)
-n^{-1}v_j^{\prime}(x) u_j(n x)
},
\quad \mu=\mu_p/\mu_{\ind{med}},
\\
&
  \label{eq:mie-bet-p}
\beta_{jm}^{(p)}=\equiv b_j^{(p)}\beta_{jm}^{(\ind{inc})}=
\frac{-i \beta_{jm}^{(\ind{inc})}}{%
n^{-1}v_j(x) u_j^{\prime}(n x)-
\mu^{-1} v_j^{\prime}(x) u_j(n x)
},
\quad
n=n_p/n_{\ind{med}},  
\end{align}
where $a_j^{(p)}$ and $b_j^{(p)}$
are the \textit{internal field coefficients};
$x=k_{\ind{med}} R_p$,
$u_j(x)= x j_j(x)$
and $v_j(x)= x h_j^{(1)}(x)$.
The similar result relating the scattered wave and the incident wave
\begin{align}
&
  \label{eq:mie-alp-sca}
  \alpha_{jm}^{(\ind{sca})}=
T_j^{11}\alpha_{jm}^{(\ind{inc})}=
\frac{n^{-1}u_j^{\prime}(x) u_j(n x)-\mu^{-1} u_j(x)
  u_j^{\prime}(n x)}{%
\mu^{-1} v_j(x) u_j^{\prime}(n x)
-n^{-1}v_j^{\prime}(x) u_j(n x)
}\alpha_{jm}^{(\ind{inc})},
\\
&
  \label{eq:mie-bet-sca}
  \beta_{jm}^{(\ind{sca})}=
T_j^{22}\beta_{jm}^{(\ind{inc})}=
\frac{
\mu^{-1} u_j(x) u_j^{\prime}(n x)-
n^{-1}u_j^{\prime}(x) u_j(n x)}{%
n^{-1}v_j(x) u_j^{\prime}(n x)-
\mu^{-1} v_j^{\prime}(x) u_j(n x)
}\beta_{jm}^{(\ind{inc})},
\end{align}
defines the \textit{T}-matrix for the simplest case of a spherically symmetric scatterer.
In addition, 
since the parity of electric and magnetic harmonics  
with respect to the spatial inversion $\uvc{r}\to -\uvc{r}$
($\{\phi,\theta\}\to \{\phi+\pi,\pi-\theta\}$)
is different
\begin{align}
  \label{eq:parity}
  \vc{Y}_{jm}^{(m)}(-\uvc{r})=(-1)^j \vc{Y}_{jm}^{(m)}(\uvc{r}),\quad
   \vc{Y}_{jm}^{(e)}(-\uvc{r})=(-1)^{j+1} \vc{Y}_{jm}^{(e)}(\uvc{r}),
\end{align}
where $f(\uvc{r})\equiv f(\phi,\theta)$ and $f(-\uvc{r})\equiv f(\phi+\pi,\pi-\theta)$,
they do not mix provided the mirror symmetry has not been broken. 
In this case the \textit{T}-matrix is diagonal and
$T_j^{12}=T_j^{21}=0$. The diagonal elements 
$T_j^{11}\equiv a_j$ and $T_j^{22}\equiv b_j$ are also called the \textit{Mie coefficients}.

\section{Far-field matching}
\label{sec:matching}

Formulas~\eqref{eq:mie-alp-p}--\eqref{eq:mie-bet-sca} 
describe the wavefield 
both inside and outside the particle when the expansion for the incident
light beam is known. 
In this section we, following Ref.~\cite{Kiselev:pra:2014}, 
apply the far-field matching method to evaluate the beam shape
coefficients. To this end, we introduce
the vector angular distributions characterizing
the wave field in the far-filed region.
The coefficients are then derived
by matching the far-field distributions for the incident 
wave and the corresponding 
expansions over vector spherical harmonics.

\subsection{Beam shape coefficients}
\label{subsec:beam-shape}

Our first step is to examine asymptotic behavior
of the total optical field, which is a sum of the incident and
scattered wave fields, 
in the far-field region, $\rho\gg 1$.
The electric and magnetic fields in this region 
can be separated into the incoming
and the outgoing parts
\begin{align}
  &
  \label{eq:E_tot-asympt}
 \vc{E}_{\ind{tot}}=\vc{E}_{\ind{inc}}+\vc{E}_{\ind{sca}}\sim
 \vc{E}_{\ind{tot}}^{(\infty)}=
\frac{1}{\rho} 
\bigl[
\exp(i\rho)
\vc{E}_{\ind{out}}(\uvc{r})
+\exp(-i\rho)
\vc{E}_{\ind{in}}(\uvc{r})
\bigr],
\\
&
  \label{eq:H_tot-asympt}
 \vc{H}_{\ind{tot}}=\vc{H}_{\ind{inc}}+\vc{H}_{\ind{sca}}\sim
\vc{H}_{\ind{tot}}^{(\infty)}=
\frac{1}{\rho}
\bigl[
\exp(i\rho)
\vc{H}_{\ind{out}}(\uvc{r})
+\exp(-i\rho)
\vc{H}_{\ind{in}}(\uvc{r})
\bigr]
\end{align}
described by 
the far-field angular
distributions: 
$\{\vc{E}_{\ind{in}},\vc{H}_{\ind{in}}\}$
and
$\{\vc{E}_{\ind{out}},\vc{H}_{\ind{out}}\}$.
These far-field vector amplitudes
are normal to $\uvc{r}$
and their basic properties can be summarized
by the following relations~\cite{Mishchenko:bk:2004}:  
\begin{align}
&
  \label{eq:H_tot_in-out}
\mu/n\,\vc{H}_{\ind{out}}(\uvc{r})=
\uvc{r}\times \vc{E}_{\ind{out}}(\uvc{r}),
\quad
\mu/n\,\vc{H}_{\ind{in}}(\uvc{r})=
-\uvc{r}\times \vc{E}_{\ind{in}}(\uvc{r}),
\\
&
\label{eq:E-in-out-tot}
\vc{E}_{\ind{out}}(\uvc{r})=\vc{E}_{\ind{out}}^{\ind{(inc)}}(\uvc{r})+\vc{E}_{\ind{out}}^{\ind{(sca)}}(\uvc{r})\perp\uvc{r},
\quad
\vc{E}_{\ind{in}}(\uvc{r})=-\vc{E}_{\ind{out}}^{\ind{(inc)}}(-\uvc{r}).
\end{align}
Formulas~\eqref{eq:E_tot-asympt}--\eqref{eq:E-in-out-tot}
explicitly show that, in the far-field region,
the incident wave field is defined by
the electric-field angular
distribution of the outgoing wave: $\vc{E}_{\ind{out}}^{\ind{(inc)}}(\uvc{r})$.
When the incident electromagnetic wave 
is represented by a  superposition of
propagating plane waves of the from
\begin{align}
&
\label{eq:E_plane-w-comb}
    \vc{E}_{\ind{inc}}(\vc{r})\equiv
  \vc{E}_{\ind{inc}}(\rho,\uvc{r})=\langle\exp(i\rho\, \uvc{k}\cdot\uvc{r})\,
  \vc{E}_{\ind{inc}}(\uvc{k})
\rangle_{\uvc{k}},
\quad
  \vc{E}_{\ind{inc}}(\uvc{k})=
\sum_{\nu=\pm 1}
E_{\nu}(\uvc{k})\,\vc{e}_{\nu}(\uvc{k}),
\end{align}
where 
$\displaystyle
\langle\,f\,\rangle_{\uvc{k}}\equiv\int_0^{2\pi}\dd\phi_k 
\int_0^{\pi}\sin\theta_k\dd\theta_k\,f$,
the distribution $\vc{E}_{\ind{out}}^{\ind{(inc)}}(\uvc{r})$
is determined by the vector amplitudes of the plane waves
as follows
\begin{align}
  \label{eq:E_out}
\vc{E}_{\ind{out}}^{\ind{(inc)}}(\uvc{r})
=-2\pi i\,
\vc{E}_{\ind{inc}}(\uvc{r})=
 E_{\theta}^{(\ind{out})}(\uvc{r})\,
\vc{e}_{\theta}(\uvc{r})
+
E_{\phi}^{(\ind{out})}(\uvc{r})\, 
\vc{e}_{\phi}(\uvc{r}),
\end{align}
whereas the incoming part of the incident wave
is described by the far-field angular distribution
$\vc{E}_{\ind{in}}^{\ind{(inc)}}(\uvc{r})=-\vc{E}_{\ind{out}}^{\ind{(inc)}}(-\uvc{r})$.

An important consequence
of Eqs.~\eqref{eq:E_plane-w-comb}
and~\eqref{eq:E_out}
is that, 
translation of the wave fields
\begin{align}
  \label{eq:shift_EH}
  \{\vc{E}_{\ind{inc}}(\vc{r}),\vc{H}_{\ind{inc}}(\vc{r})\}
\to
\{\vc{E}_{\ind{inc}}(\vc{r}+\vc{r}_p),\vc{H}_{\ind{inc}}(\vc{r}+\vc{r}_p)\}
\end{align}
affects the far-field angular distribution~\eqref{eq:E_out}
by producing the phase shift
\begin{align}
  \label{eq:shift_E_out}
  \vc{E}_{\ind{out}}^{\ind{(inc)}}(\uvc{r})
\to
\vc{E}_{\ind{out}}^{\ind{(inc)}}(\uvc{r},\vc{r}_p)=
\vc{E}_{\ind{out}}^{\ind{(inc)}}(\uvc{r})\exp[i k (\vc{r}_p\cdot\uvc{r})].
\end{align}
Referring to Fig.~\ref{fig:beam},
the vector $-\vc{r}_p$
will determine location of
the beam waist with respect to the center
of the particle.
 
The far-field distribution
of an incident light beam,
$\vc{E}_{\ind{out}}^{\ind{(inc)}}(\uvc{r})$,
can also be found from the expansion
over the vector spherical harmonics~\eqref{eq:E_alpha}.  
The far-field asymptotics
for the vector wave functions
that enter the expansion for the
incident wave~\eqref{eq:EH}
\begin{align}
&
  \label{eq:M_inc-asympt}
 \vc{M}_{jm}^{(\ind{inc})}(\rho,\uvc{r})\sim
\frac{(-i)^{j+1}}{2 \rho}
\bigl[
\exp(i\rho)
\vc{Y}_{jm}^{(m)}(\uvc{r})
-\exp(-i\rho)
\vc{Y}_{jm}^{(m)}(-\uvc{r})
\bigr],
\\
&
  \label{eq:N_inc-asympt}
 \vc{N}_{jm}^{(\ind{inc})}(\rho,\uvc{r})\sim
\frac{(-i)^j}{2 \rho}
\bigl[
\exp(i\rho)
\vc{Y}_{jm}^{(e)}(\uvc{r})
-\exp(-i\rho)
\vc{Y}_{jm}^{(e)}(-\uvc{r})
\bigr],
\end{align}
can be derived from Eqs.~\eqref{eq:M_jm}-\eqref{eq:N_jm}
with the help of the far-field relation~\eqref{eq:asymp-bessel}.
Substituting Eqs.~\eqref{eq:M_inc-asympt} and~\eqref{eq:N_inc-asympt}
into Eq.~\eqref{eq:E_alpha} gives
the expansion for 
the far-field distribution~\eqref{eq:E_out}
\begin{align}
  \label{eq:E_out_expan}
 \vc{E}_{\ind{out}}^{\ind{(inc)}}(\uvc{r})=
2^{-1}
\sum_{jm} 
(-i)^{j+1} \Bigl[
 \alpha_{jm}^{(\ind{inc})}
\vc{Y}_{jm}^{(m)}(\uvc{r})
+
i \beta_{jm}^{(\ind{inc})}
\vc{Y}_{jm}^{(e)}(\uvc{r})
\Bigr]\equiv
\sum_{jm}\sum_{\alpha\in\{e,m\}} w_{jm}^{(\alpha)}\vc{Y}_{jm}^{(\alpha)},
\end{align}
where $w_{jm}^{(m)}=(-i)^{j+1}\alpha_{jm}^{(\ind{inc})}/2$
and $w_{jm}^{(e)}=(-i)^{j}\beta_{jm}^{(\ind{inc})}/2$.
Similar result
for
the far-field angular distribution
of the scattered wave, 
$\vc{E}_{\ind{out}}^{\ind{(sca)}}(\uvc{r})$,
is given by
\begin{align}
  \label{eq:E_out_sca_expan}
 \vc{E}_{\ind{out}}^{\ind{(sca)}}(\uvc{r})=
\sum_{jm} 
(-i)^{j+1} \Bigl[
 \alpha_{jm}^{(\ind{sca})}
\vc{Y}_{jm}^{(m)}(\uvc{r})
+
i \beta_{jm}^{(\ind{sca})}
\vc{Y}_{jm}^{(e)}(\uvc{r})
\Bigr]
\equiv
\sum_{jm}\sum_{\alpha\in\{e,m\}} s_{jm}^{(\alpha)}\vc{Y}_{jm}^{(\alpha)},
\end{align}
where $s_{jm}^{(m)}=(-i)^{j+1}\alpha_{jm}^{(\ind{sca})}$
and $s_{jm}^{(e)}=(-i)^{j}\beta_{jm}^{(\ind{sca})}$.

The coefficients of the incident wave
can now be easily found as the Fourier coefficients 
of the far-field angular distribution, $\vc{E}_{\ind{out}}^{(\ind{inc})}$,
expanded using the vector spherical harmonics basis~\eqref{eq:Y_D}.
The final result reads
\begin{subequations}
\label{eq:alp_beta_inc2}
\begin{align}
&
  \label{eq:alp_inc2}
\alpha_{jm}^{(\ind{inc})}= 
2\, i^{j+1}
  \langle \vc{Y}_{jm}^{(m)\,*}(\uvc{r})\cdot
\vc{E}_{\ind{out}}^{\ind{(inc)}}(\uvc{r})
\rangle_{\uvc{r}}=
2 n_j\, i^{j+1}
  \langle {Y}_{jm}^{\,*}(\uvc{r})\,
(\vc{L}
\cdot
\vc{E}_{\ind{out}}^{\ind{(inc)}}(\uvc{r}))
\rangle_{\uvc{r}}=
\notag
\\
&
2 n_j\, i^{j}
\int_{0}^{2\pi}\dd\phi\int_{0}^{\pi}\dd\theta\,
{Y}_{jm}^{\,*}(\phi,\theta)
\Bigl[
\prt{\theta}(\sin\theta E_{\phi}^{(\ind{out})})
- \prt{\phi}E_{\theta}^{(\ind{out})}
\Bigr],
\\
&
  \label{eq:beta_inc2}
 \beta_{jm}^{(\ind{inc})}= 
2\, i^{j}
  \langle \vc{Y}_{jm}^{(e)\,*}(\uvc{r})\cdot
\vc{E}_{\ind{out}}^{\ind{(inc)}}(\uvc{r})
\rangle_{\uvc{r}}=
-2 n_j\, i^{j}\,
  \langle {Y}_{jm}^{\,*}(\uvc{r})\,
(r \bs{\nabla}
\cdot
\vc{E}_{\ind{out}}^{\ind{(inc)}}(\uvc{r}))
\rangle_{\uvc{r}}=
\notag
\\
&
-2 n_j\, i^{j}
\int_{0}^{2\pi}\dd\phi\int_{0}^{\pi}\dd\theta\,
{Y}_{jm}^{\,*}(\phi,\theta)
\Bigl[
\prt{\theta}(\sin\theta E_{\theta}^{(\ind{out})})
+ \prt{\phi}E_{\phi}^{(\ind{out})}
\Bigr],
\end{align}
\end{subequations}
where we have used
Eqs.~\eqref{eq:Ym-in-Yjm} and~\eqref{eq:Ye-in-Yjm}
to obtain the explicit analytical
expressions
useful for computational purposes.

\subsection{Remodelled Laguerre--Gaussian beams}
\label{subsec:LG-beams}


In the paraxial approximation,
the LG beams are described in terms of
scalar fields of the form: $u_{nm}(\vc{r}) \exp(i k z)$,
where $n$ ($m$) is the radial (azimuthal) mode number
and
$u_{nm}(\vc{r})$ is the solution
of the paraxial Helmholtz equation
that can be conveniently
written in the cylindrical coordinate system,
$(r_{\perp},\phi,z)$, as follows
\begin{subequations}
  \label{eq:LG-scalar}
\begin{align}
&
  \label{eq:LG-u_mn}
  u_{nm}(r_{\perp},\phi,z)=
  |\sigma|^{-1}\psi_{nm}(\sqrt{2}r_{\perp}/w)
\exp\{
-r_{\perp}^2/(w_0^2\sigma)
+ i m \phi - i \gamma_{nm}
\},
\\
&
  \label{eq:LG-sigma-w}
\sigma\equiv\sigma(z)=1+i z/z_R,
\quad
w\equiv w(z)=w_0|\sigma|,
\\
&
  \label{eq:LG-gamma-psi}
\gamma_{nm}\equiv\gamma_{nm}(z)=(2n+m+1)\arctan(z/z_R),
\quad
\psi_{nm}(x)=x^{|m|} L_n^{|m|}(x^2),
\end{align}
\end{subequations}
where
$L_n^{m}$ is the generalized Laguerre polynomial given by~\cite{Grad:1980}
\begin{align}
  \label{eq:L_nm}
  L_n^m(x)=(n!)^{-1} x^{-m} \exp(x)\,  \prt{x}^{n}\,[x^{n+m}\exp(-x)],
\end{align}
$w_0$ is the initial transverse Gaussian half-width
(the beam diameter at waist)
$z_R=k w_0^2/2=[2 k f^2]^{-1}$ is the Rayleigh range
and $f=[k w_0]^{-1}$ is the \textit{focusing parameter}.

The problem studied in Refs.~\cite{Zhou:ol:2006,Duan:josaa:2005,Zhou:olt:2008}
deals with  the exact propagation of the optical
field in the half-space, $z>0$, when 
its transverse components at the initial (source) plane,
$z=0$, are known.
In Ref.~\cite{Zhou:ol:2006}, 
the results describing asymptotic behavior of the linearly polarized 
field
\begin{align}
  \label{eq:LG-init}
  \vc{E}(r_{\perp},\phi,0)=u_{nm}(r_{\perp},\phi,0)\,\uvc{x}=
  \psi_{nm}(\sqrt{2}r_{\perp}/w_0)
\exp\{
-r_{\perp}^2/w_0^2
+ i m \phi\}\,\uvc{x}
\end{align}
were derived using the angular spectrum representation
(Debye integrals)
and comply with both
the results of rigorous mathematical analysis
performed in Ref.~\cite{Sherman:jmath:1976}
and those obtained using the vectorial Rayleigh-Sommerfeld
integrals~\cite{Duan:josaa:2005,Zhou:olt:2008}.
The resulting expression for the far-field angular distribution
can be written in the following form
\begin{subequations}
  \label{eq:E_ff_LG}
\begin{align}
&
  \label{eq:E_out_LG}
  \vc{E}_{\ind{out}}^{(\ind{LG})}(\phi,\theta)=E_{nm}(f^{-1}\sin\theta/\sqrt{2})\,
\exp(im \phi)
\vc{e}_{\ind{out}},
\\
&
 \label{eq:e_out}
\vc{e}_{\ind{out}}
=
\cos\phi\,\vc{e}_{\theta}(\uvc{r})-
\cos\theta\,\sin\phi\,\vc{e}_{\phi}(\uvc{r})
=
\cos\theta\,\uvc{x}-\sin\theta\,\cos\phi\,\uvc{z},
\\
&
 \label{eq:E_nm}
E_{nm}(x)=\frac{ x^m}{i^{2n+m+1} 2 f^{2}}\,
L_n^{m}(x^2)\,\exp(-x^2/2).
\end{align}
\end{subequations}

The beam shape coefficients for
the LG beams can now be computed from
formulas~\eqref{eq:alp_beta_inc2}
where the far-field distribution
$\vc{E}_{\ind{out}}^{(\ind{inc})}$
is defined in Eq.~\eqref{eq:E_out_LG}.
We can also combine the relations~\eqref{eq:E_out}
and~\eqref{eq:E_plane-w-comb}
with the outgoing part of
the far-field distribution~\eqref{eq:E_out_LG}
to deduce the expression for the electric field
of the remodelled LG beam
\begin{align}
&
  \label{eq:E_inc_LG}
  \vc{E}_{\ind{inc}}^{(\ind{LG})}(\rho_\perp,\phi,\rho_z)=
  E_{x}^{(\ind{LG})}(\rho_\perp,\phi,\rho_z)\,\uvc{x}+
  E_{z}^{(\ind{LG})}(\rho_\perp,\phi,\rho_z)\,\uvc{z}=
\notag
\\
&
\frac{i}{2\pi}
\langle
\exp\left[
i(
\rho_\perp\sin\theta_k\cos(\phi-\phi_k)
+\rho_z\cos\theta_k
)
\right]
\,
\vc{E}_{\ind{out}}^{(\ind{LG})}(\uvc{k})
\rangle_{\uvc{k}},
\end{align}
where $\rho_\perp=k r_\perp$ and $\rho_z=k z$.

\section{Optical force and symmetries}
\label{subsec:opt-force}

The electric-field far-field distributions for the incident and the scattered waves
(see Eqs~\eqref{eq:E_out_expan}
and~\eqref{eq:E_out_sca_expan}, respectively)
are found to 
play a major part in the method of far-field matching.
In particular,
they determine the beam shape coefficients~\eqref{eq:alp_beta_inc2}
and incorporate dependence on the particle position
[see Eq.~\eqref{eq:shift_E_out}].
In this section, 
we derive a useful 
far-field expression for the optical force
and discuss some symmetry properties
of the LG beams.

\subsection{Maxwell's stress tensor and optical force}
\label{subsec:ff-poynting}

It is not difficult to obtain the far-field expression for 
the time-averaged Poynting vector of 
the total wavefield given in
Eqs.~\eqref{eq:E_tot-asympt}--\eqref{eq:H_tot_in-out}
$\vc{S}_{\ind{tot}}=c/(8\pi)\Re(\vc{E}_{\ind{tot}}\times\vc{H}_{\ind{tot}}^{\,*})$
\begin{align}
&
  \label{eq:S_tot-asympt}
\vc{S}_{\ind{tot}}(\rho,\uvc{r})\sim
\vc{S}_{\ind{tot}}^{(\infty)}(\rho,\uvc{r})=
\frac{c n}{8\pi\mu\rho^2}
\Bigl\{
|\vc{E}_{\ind{out}}(\uvc{r})|^2
-
|\vc{E}_{\ind{in}}(\uvc{r})|^2
\Bigr\}
\uvc{r}
\end{align}
where
$|\vc{E}_{\alpha}(\uvc{r})|^2=(\vc{E}_{\alpha}(\uvc{r})\cdot\vc{E}_{\alpha}^{\,*}(\uvc{r}))$,
and use the relations~\eqref{eq:E-in-out-tot}
to evaluate the flux of the Poynting vector~\eqref{eq:S_tot-asympt}
through the far-field sphere $S_f$ of the radius $R_f$.

The result can be written in the following well-known form:
\begin{align}
&
  \label{eq:S_tot_flux}
\oiint\limits_{S_f}(\vc{S}_{\ind{tot}}^{(\infty)}\cdot\dd\vc{s})
=
  R_f^2 
\langle
  (\vc{S}_{\ind{tot}}^{(\infty)}(kR_f,\uvc{r})\cdot\uvc{r})
  \rangle_{\uvc{r}}
\equiv -W_{\ind{abs}}=W_{\ind{sca}}-W_{\ind{ext}},
\\
&
\label{eq:W_sca_ext}
 W_{\ind{sca}}=\frac{c n}{8\pi \mu k^2}
\langle
|\vc{E}_{\ind{out}}^{(\ind{sca})}(\uvc{r})|^2
  \rangle_{\uvc{r}},
\quad
 W_{\ind{ext}}=-\frac{c n}{4\pi \mu k^2}
\Re
\langle
(\vc{E}_{\ind{out}}^{\ind{(sca)}}(\uvc{r})\cdot[\vc{E}_{\ind{out}}^{\ind{(inc)}}(\uvc{r})]^{\,*})
  \rangle_{\uvc{r}},
\end{align}
where $W_{\ind{sca}}$ is the energy scattering rate
(the rate at which the scattered energy crosses the sphere in the
outward direction),
$W_{\ind{abs}}$ is the energy absorption rate
and $W_{\ind{ext}}=W_{\ind{sca}}+W_{\ind{abs}}$ is the extinction
rate. When the scatterer and the surrounding medium
are both non-absorbing, the energy absorption rate vanishes,
$W_{\ind{abs}}=0$, and Eq.~\eqref{eq:S_tot_flux}
yields unitarity relations for the
\textit{T}-matrix~\cite{Mishchenko:bk:2004}
(see also Eq.~\eqref{eq:T-unit-oprt} in Appendix~\ref{sec:derivation}).
In our spherically symmetric case,
these are: $|2T_j^{11}+1|^2=|2T_j^{22}+1|^2=1$.

According to Ref.~\cite{Simpson:josaa:1:2009},
the total power of the incident laser beam,
$W_{\ind{inc}}$, can be computed 
by integrating the Poynting vector for
the outgoing part of the incident field.
In our case, this part
expressed in terms of the beam shape coefficients 
is given in Eq.~\eqref{eq:E_out_expan}
and the total power of the incident beam
can be written in the form of a sum:
\begin{align}
  \label{eq:W_inc}
  W_{\ind{inc}}=\frac{c n}{8\pi \mu k^2}
\langle
|\vc{E}_{\ind{out}}^{(\ind{inc})}(\uvc{r})|^2
  \rangle_{\uvc{r}}
= 
\sum_j\bigl\{
W_j^{(m)}+W_j^{(e)}
\bigr\},
\end{align}
where 
$W_j^{(m)}$ ($W_j^{(e)}$)
is the power of magnetic (electric)
modes with the angular momentum $j$
given by
\begin{align}
  \label{eq:W_j}
  W_j^{(m)}=
\frac{c n}{32\pi \mu k^2}
\sum_{m=-j}^{j}
|\alpha_{jm}^{(\ind{inc})}|^2,
\quad
  W_j^{(e)}=
\frac{c n}{32\pi \mu k^2}
\sum_{m=-j}^{j}
|\beta_{jm}^{(\ind{inc})}|^2.
\end{align}

The far-field angular distributions,
$\vc{E}_{\ind{out}}^{\ind{(sca)}}(\uvc{r})$
and
$\vc{E}_{\ind{out}}^{\ind{(inc)}}(\uvc{r})$,
also determine
the time-averaged optical force, $\vc{F}$, acting upon the 
particle. This force can be expressed
in terms of the time-average of Maxwell's stress tensor
$\mvc{T}_{M}$ 
\begin{align}
  \label{eq:T_M_gen}
  \mvc{T}_M=
\frac{1}{8\pi}
\Re
\{
\epsilon \vc{E}\otimes\vc{E}^{*}+
\mu \vc{H}\otimes\vc{H}^{*}
-\mvc{I}
(
\epsilon |\vc{E}|^2
+\mu |\vc{H}|^2
)/2
\},
\end{align}
where $\mvc{I}$ is the unit dyadic,
as follows:
\begin{align}
  \label{eq:F_gen}
  \vc{F}=\oiint\limits_{S_f}(\mvc{T}_{M}^{(\infty)}\cdot\dd\vc{s}),
\end{align}
where $\mvc{T}_{M}^{(\infty)}$ is 
the Maxwell stress tensor~\eqref{eq:T_M_gen}
in the far-field region.
Substituting Eqs.~\eqref{eq:E_tot-asympt}--\eqref{eq:H_tot_in-out}
into the stress tensor~\eqref{eq:T_M_gen}
gives the following expression for
the dot product
\begin{align}
  \label{eq:sca_T_M-ff}
  (\mvc{T}_{M}^{(\infty)}\cdot\uvc{r})=
-\frac{\epsilon}{8\pi\rho^2}
\Bigl\{
|\vc{E}_{\ind{out}}(\uvc{r})|^2
+
|\vc{E}_{\ind{in}}(\uvc{r})|^2
\Bigr\}
\uvc{r}
\end{align}
that enter the integrand on
the right-hand side of Eq.~\eqref{eq:F_gen}. 
The resulting expression for
the optical force is
\begin{align}
  \label{eq:F_far-field}
  \vc{F}(\vc{r}_p)=
-\frac{\epsilon}{8\pi k^2}
\Bigl\{
\langle
\uvc{r} 
|\vc{E}_{\ind{out}}^{(\ind{sca})}(\uvc{r},\vc{r}_p)|^2
\rangle_{\uvc{r}}
+
2
\Re
\langle
\uvc{r} 
([\vc{E}_{\ind{out}}^{\ind{(inc)}}(\uvc{r},\vc{r}_p)]^{\,*}\cdot\vc{E}_{\ind{out}}^{\ind{(sca)}}(\uvc{r},\vc{r}_p))
\rangle_{\uvc{r}}
\Bigr\}
,
\end{align}
where we have indicated that the net force 
exerted on the particle depends on 
the displacement vector $\vc{r}_p$ describing
position of the scatterer with respect to the focal
plane (see Fig.~\ref{fig:beam}).

In the special case of plane-wave illumination,
the far-field expression for the optical force
was derived in Ref.~\cite{Mishchenko:jqsrt:2001}. 
In Appendix~\ref{sec:derivation} we show that
formula~\eqref{eq:F_far-field}
can alternatively be recast into
the form (see Eq.~\eqref{eq:F-grad})
where the terms containing derivatives
with respect to coordinates of the displacement
vector $\vc{r}_p$ (the gradient terms) 
are explicitly separated out.

In the spherical basis,
$\uvc{e}_{\pm}=\mp(\uvc{x}\pm i\uvc{y})/\sqrt{2}$
and $\uvc{e}_0=\uvc{z}$,
the components of the optical force
can be expressed in terms of
the coefficients that enter
the expansions for the incident and scattered
waves [see Eqs.~\eqref{eq:E_out_expan} and~\eqref{eq:E_out_sca_expan}]
as follows
\begin{align}
&
  \label{eq:F_nu}
  F_\nu=(\vc{F}\cdot\uvc{e}_\nu^{\,*})=
-\frac{\epsilon}{8\pi k^2}
\sum_{j m}\sum_{j' m'}
\Bigl\{
p_{m m'}^{j\, j'} P_{m m' \nu}^{j\, j'\, 1}
+q_{m m'}^{j\, j'} Q_{m m' \nu}^{j\, j'\, 1}
\Bigr\}
\\
&
\label{eq:p_jjmm}
p_{m m'}^{j\, j'}=
\sum_{\alpha}
\Bigl\{
s_{jm}^{(\alpha)}s_{j'm'}^{(\alpha)\,*}+
s_{jm}^{(\alpha)}w_{j'm'}^{(\alpha)\,*}+
w_{jm}^{(\alpha)}s_{j'm'}^{(\alpha)\,*}
\Bigr\}
\\
&
\label{eq:q_jjmm}
q_{m m'}^{j\, j'}=
\sum_{\alpha,\,\beta}
(1-\delta_{\alpha\beta})
\Bigl\{
s_{jm}^{(\alpha)}s_{j'm'}^{(\beta)\,*}+
s_{jm}^{(\alpha)}w_{j'm'}^{(\beta)\,*}+
w_{jm}^{(\alpha)}s_{j'm'}^{(\beta)\,*}
\Bigr\},
\\
&
\label{eq:P_jjmm}
P_{m m' \nu}^{j\, j'\, 1}=
N_jN_{j'}/2
\sum_{\mu=\pm 1}
\langle
D_{m\mu}^{j\,*}(\uvc{r})D_{m'\mu}^{j'}(\uvc{r})D_{\nu 0}^{1}(\uvc{r})
\rangle_{\uvc{r}}
\notag
\\
&
=
\frac{1}{8}\sqrt{\frac{2j'+1}{2j+1}}
C_{\nu m' m}^{1\: j'\: j}\sum_{\mu=\pm 1}C_{0 \mu \mu}^{1  j' j},
\\
&
\label{eq:Q_jjmm}
Q_{m m' \nu}^{j\, j'\, 1}=
-N_jN_{j'}/2
\sum_{\mu=\pm 1}
\mu
\langle
D_{m\mu}^{j\,*}(\uvc{r})D_{m'\mu}^{j'}(\uvc{r})D_{\nu 0}^{1}(\uvc{r})
\rangle_{\uvc{r}}
\notag
\\
&
=
-\frac{1}{8}\sqrt{\frac{2j'+1}{2j+1}}
C_{\nu m' m}^{1\: j'\: j}\sum_{\mu=\pm 1}\mu C_{0 \mu \mu}^{1  j' j},
\end{align}
where $\nu\in\{\pm 1, 0\}$
and $C_{\nu m' m}^{1\: j'\: j}$ denotes 
the Clebsch-Gordon (Wigner)
coefficient.
Derivation of formula~\eqref{eq:F_nu}
involves the following steps:
(a)~substituting expansions~\eqref{eq:E_out_expan}
and~\eqref{eq:E_out_sca_expan}
into the expression for the optical force;
(b)~using the components of the vector $\uvc{r}$
expressed in terms of $D$ functions: 
$(\uvc{r}\cdot\uvc{e}_{\nu}^{*})=D_{\nu 0}^{1}(\uvc{r})$;
(c) using Eq.~\eqref{eq:Y_D} to compute
dot products of the vector spherical functions
$(\vc{Y}_{jm}^{(\alpha)}\cdot\vc{Y}_{j'm'}^{(\beta)\,*})$
and (d)~using the relation~\cite{Biedenharn:bk:1981}
\begin{align}
  \label{eq:D3-rel}
  \langle
D_{m\mu}^{j\,*}(\uvc{r})D_{m'\mu}^{j'}(\uvc{r})D_{\nu 0}^{1}(\uvc{r})
\rangle_{\uvc{r}}=
\frac{\pi}{2j+1}
C_{\nu m' m}^{1\: j'\: j}C_{0 \mu \mu}^{1  j' j}
\end{align}
to perform the integrals.

The result~\eqref{eq:F_nu} can be further simplified
by using the permutation symmetry relations
\begin{align}
  \label{eq:permut-pq}
  p_{m' m}^{j'\, j}=[p_{m m'}^{j\, j'}]^{*},
\quad
  q_{m' m}^{j'\, j}=[q_{m m'}^{j\, j'}]^{*},
\\
  \label{eq:permut-PQ}
P_{m' m \nu}^{j'\, j\, 1}=
(-1)^{\nu} P_{m m' -\nu}^{j\, j'\, 1},
\quad
Q_{m' m \nu}^{j'\, j\, 1}=
(-1)^{\nu} Q_{m m' -\nu}^{j\, j'\, 1}
\end{align}
and 
the explicit expressions for the
coefficients 
\begin{align}
&
  \label{eq:PQ_jj}
  P_{m m' \nu}^{j\, j\, 1}=0,
\quad
Q_{m m' \nu}^{j\, j\, 1}=-\frac{\delta_{m',\,m-\nu}}{4j(j+1)}
\begin{cases}
  m,&\nu=0\\
[(j+\nu m)(j-\nu m +1)/2]^{1/2},& \nu=\pm 1
\end{cases}
\\
&
  \label{eq:PQ_j-1j}
  Q_{m m' \nu}^{j-1\, j\, 1}=0,
\quad
P_{m m' \nu}^{j-1\, j\, 1}=\frac{\delta_{m',\,m-\nu}}{4j}
\sqrt{\frac{j^2-1}{4j^2-1}}
\begin{cases}
  [j^2-m^2]^{1/2},&\nu=0\\
[(j-\nu m)(j-\nu m +1)/2]^{1/2},& \nu=\pm 1
\end{cases}
\end{align}
derived with the help of
formulas for
the Clebsch-Gordon 
coefficients (see, e.g., the table on
pg.~635 of the book~\cite{Biedenharn:bk:1981}).

The final result for the components of the optical
force~\eqref{eq:F_far-field} reads
\begin{align}
&
  \label{eq:force-final}
  F_\nu=
-\frac{\epsilon}{8\pi k^2}
\sum_{j m}
\Bigl\{
q_{m m-\nu}^{j\, j} Q_{m m-\nu \nu}^{j\, j\, 1}
+
[q_{m m-\nu}^{j\, j}]^{*} (-1)^{\nu} Q_{m m+\nu -\nu}^{j\, j\, 1}
\notag
\\
&
+p_{m m-\nu}^{j-1\, j} P_{m m-\nu \nu}^{j-1\, j\, 1}
+
[p_{m m-\nu}^{j-1\, j}]^{*} (-1)^{\nu} P_{m m+\nu -\nu}^{j-1\, j\, 1}
\Bigr\}.
\end{align}
Note that it is often useful to rescale the force
by introducing  
the dimensionless force efficiency~\cite{Nieminen:jqsrt:2014}
\begin{align}
  \label{eq:force-efficiency}
  \vc{F}_{\ind{eff}}=\vc{F}/F_{\ind{scl}},
\quad
F_{\ind{scl}}=n W_{\ind{inc}}/c,
\end{align}
where $F_{\ind{scl}}$ is the force scale factor
proportional to the power of the incident
beam $W_{\ind{inc}}$ given by Eq.~\eqref{eq:W_inc}.

\subsection{Symmetries of laser beams and stiffness matrix}
\label{subsec:symmetry}

In Sec.~\ref{subsec:ff-poynting},
we have shown that the
scattering characteristics
such as the cross-sections and the radiation force
can be expressed in terms of the
far-field angular distributions
that can be regarded as vector fields on a sphere.
Under the action of the orthogonal transformation
$M$: $\uvc{r}\mapsto \uvc{r}'=M\uvc{r}$
 such fields transform as follows: 

\begin{align}
  \label{eq:E_out-transf-gen}
  \vc{E}_{\ind{out}}(\uvc{r})\mapsto \vc{E}_{\ind{out}}^{\,\prime}=
M\vc{E}_{\ind{out}}(M^{-1}\uvc{r}).
\end{align}
From Eqs.~\eqref{eq:E_plane-w-comb} and~\eqref{eq:F_far-field}
we derive the relations
\begin{align}
  \label{eq:E_inc-transf-gen}
  \vc{E}_{\ind{inc}}(\vc{r})\mapsto \vc{E}_{\ind{inc}}^{\,\prime}=
M\vc{E}_{\ind{inc}}(M^{-1}\vc{r}),
\quad
  \vc{F}[\vc{E}_{\ind{out}}]\mapsto \vc{F}[\vc{E}_{\ind{out}}^{\,\prime}]=
M\vc{F}[\vc{E}_{\ind{out}}]
\end{align}
that define transformations of the incident wave and the optical
force.

The symmetry transformation $M_s$
for the far-field angular distribution
of the incident wave may generally be defined through 
the symmetry relation
\begin{align}
  \label{eq:E_out-transf-symmetry}
  M_s\vc{E}_{\ind{out}}^{(\ind{inc})}(M_s^{-1}\uvc{r})=
p_s\vc{E}_{\ind{out}}^{(\ind{inc})}(\uvc{r}),
\end{align}
where $p_s\equiv\exp(i\psi_s)$ is the phase factor.
At $|\vc{r}_p|\ne 0$,
we can use Eq.~\eqref{eq:shift_E_out}
combined with
the orthogonality relation:
$(\uvc{r}\cdot\vc{r}_p)=(M_s^{-1}\uvc{r}\cdot M_s^{-1}\vc{r}_p)$
to recast
the symmetry condition~\eqref{eq:E_out-transf-symmetry} 
in the form:
\begin{align}
  \label{eq:E_out-symmetry-Rd}
p_s\vc{E}_{\ind{out}}^{(\ind{inc})}(\uvc{r},\vc{r}_p)=
  M_s\vc{E}_{\ind{out}}^{(\ind{inc})}(M_s^{-1}\uvc{r},M_s^{-1}\vc{r}_p).
\end{align}
As a direct consequence
of the generalized symmetry relation~\eqref{eq:E_out-symmetry-Rd} 
for the optical force we have
\begin{align}
&
  \label{eq:KF-symmetry}
  \vc{F}(\vc{r}_p)=M_s\vc{F}(M_s^{-1}\vc{r}_p),
\quad
 \mvc{K}(\vc{r}_p)=M_s\mvc{K}(M_s^{-1}\vc{r}_p)M_s^{-1},
\end{align}
where the elements of the stiffness (force) matrix
$\mvc{K}(\vc{r}_p)$ are given by
\begin{align}
  \label{eq:K-def}
K_{ij}(\vc{r}_p)=\prt{j}F_i(\vc{r}_p).  
\end{align}
At equilibria, the force vanishes ($\vc{F}(\vc{r}_{\ind{eq}})=\vc{0}$) 
and the stiffness matrix,
$\mvc{K}_{\ind{eq}}=\mvc{K}(\vc{r}_{\ind{eq}})$,
is known to
govern the regime of 
linearized dynamics of the particle~\cite{Simpson:pre:2010}. 

For the LG beams 
with the angular distribution~\eqref{eq:E_out_LG},
it can be easily checked that
the direction of propagation 
(the $z$ axis) is 
the axis of twofold rotational symmetry $C_2$
with
$C_2:\phi\mapsto \phi+\pi$ and $C_2=\diag(-1,-1,1)$.
From Eq.~\eqref{eq:E_out_LG}, we have
\begin{align}
  \label{eq:LG-C2}
  C_2 \vc{E}_{\ind{out}}^{(\ind{LG})}(C_2 \uvc{r})
=C_2 \vc{E}_{\ind{out}}^{(\ind{LG})}(\phi+\pi,\theta)=
(-1)^{m+1}\vc{E}_{\ind{out}}^{(\ind{LG})}(\uvc{r}).
\end{align}

When $\vc{r}_p\parallel\uvc{z}$ and $C_2\vc{r}_p=\vc{r}_p$,
equation~\eqref{eq:KF-symmetry} for the twofold symmetry 
implies
that the optical force is directed along the symmetry axis,
$\vc{F}\parallel\uvc{z}$, and the stiffness matrix
is of the form:
\begin{align}
  \label{eq:K-C2}
  \mvc{K}=
  \begin{pmatrix}
    K_{xx} & K_{xy} & 0\\
K_{yx} & K_{yy} & 0\\
0 & 0 & K_{zz}
  \end{pmatrix}.
\end{align}
Since
$C_2\vc{Y}_{jm}^{(e,\,m)}(C_2\uvc{r})=(-1)^m\vc{Y}_{jm}^{(e,\,m)}(\uvc{r})$,
for $C_2$ symmetric LG beams,
the azimuthal numbers of nonvanishing beam shape coefficients
are of the same parity (all $m$ are either odd or even). 

We conclude this section with the remark on
the special case of non-vortex LG beams  
with the vanishing azimuthal mode number. 
At $m=0$,
the angular distribution~\eqref{eq:E_out_LG}
is invariant under the reflection
\begin{align}
  \label{eq:LG-sigma}
  \sigma_{xz} \vc{E}_{\ind{out}}^{(\ind{LG})}(\sigma_{xz} \uvc{r})
=\sigma_{xz} \vc{E}_{\ind{out}}^{(\ind{LG})}(-\phi,\theta)=
\vc{E}_{\ind{out}}^{(\ind{LG})}(\uvc{r}),
\end{align}
 where $\sigma_{xz}=\diag(1,-1,1)$.
This mirror plane symmetry
places additional constraints on 
the elements of the
stiffness matrix
at $\vc{r}_d=\sigma_{xz}\vc{r}_d$.
From Eq.~\eqref{eq:KF-symmetry},
it can be inferred that  
the non-diagonal elements
$K_{xy}$ and $K_{yx}$ should be equal to zero.
So, for non-vortex beams with $m=0$,
the matrix~\eqref{eq:K-C2}  
is diagonal
\begin{align}
  \label{eq:K-m0}
  \mvc{K}=\diag(K_{xx},K_{yy},K_{zz}).
\end{align}

\section{Results}
\label{sec:results}

In this section,
we 
present the results of numerical 
computations on the radiation force~\eqref{eq:F_far-field}
for the case
where
the incident wave 
is represented by 
the remodelled LG beams~\eqref{eq:E_inc_LG}
with the radial mode number $n=n_{\ind{LG}}\in\{0,1\}$ and
the azimuthal number, $m=m_{\ind{LG}}\in\{0,1,2\}$.
Substituting the far-field distribution~\eqref{eq:E_ff_LG}
into Eq.~\eqref{eq:alp_beta_inc2}
gives the beam shape coefficients of these beams
in the form that agrees with our symmetry analysis:
\begin{subequations}
\label{eq:alpha-beta-inc-rules}
\begin{align}
&
  \label{eq:alpha-inc-rules}
  \alpha_{jm}^{(\ind{inc})}=
\alpha_{j,\,m_{\ind{LG}}}^{(+)}\,\delta_{m,\,m_{\ind{LG}}+1}+
\alpha_{j,\,m_{\ind{LG}}}^{(-)}\,\delta_{m,\,m_{\ind{LG}}-1},
\\
&
\label{eq:beta-inc-rules}
  \beta_{jm}^{(\ind{\ind{inc}})}=
\beta_{j,\,m_{\ind{LG}}}^{(+)}\,\delta_{m,\,m_{\ind{LG}}+1}+
\beta_{j,\,m_{\ind{LG}}}^{(-)}\,\delta_{m,\,m_{\ind{LG}}-1}.
\end{align}
\end{subequations}
Then 
the coefficients of expansions~\eqref{eq:EH}
describing scattered wave
and electromagnetic field inside the scatterer
can be evaluated from 
formulas~\eqref{eq:mie-alp-p}--\eqref{eq:mie-bet-sca}.
These coefficients enter the expression for 
the components of
the optical force~\eqref{eq:force-final}.
The optical-force-induced dynamics of the particle
will be of our primary concern.

\subsection{Linearized dynamics and stability of equilibria}
\label{subsec:dynamics}

\begin{figure*}[!tbh]
\includegraphics[width=120mm]{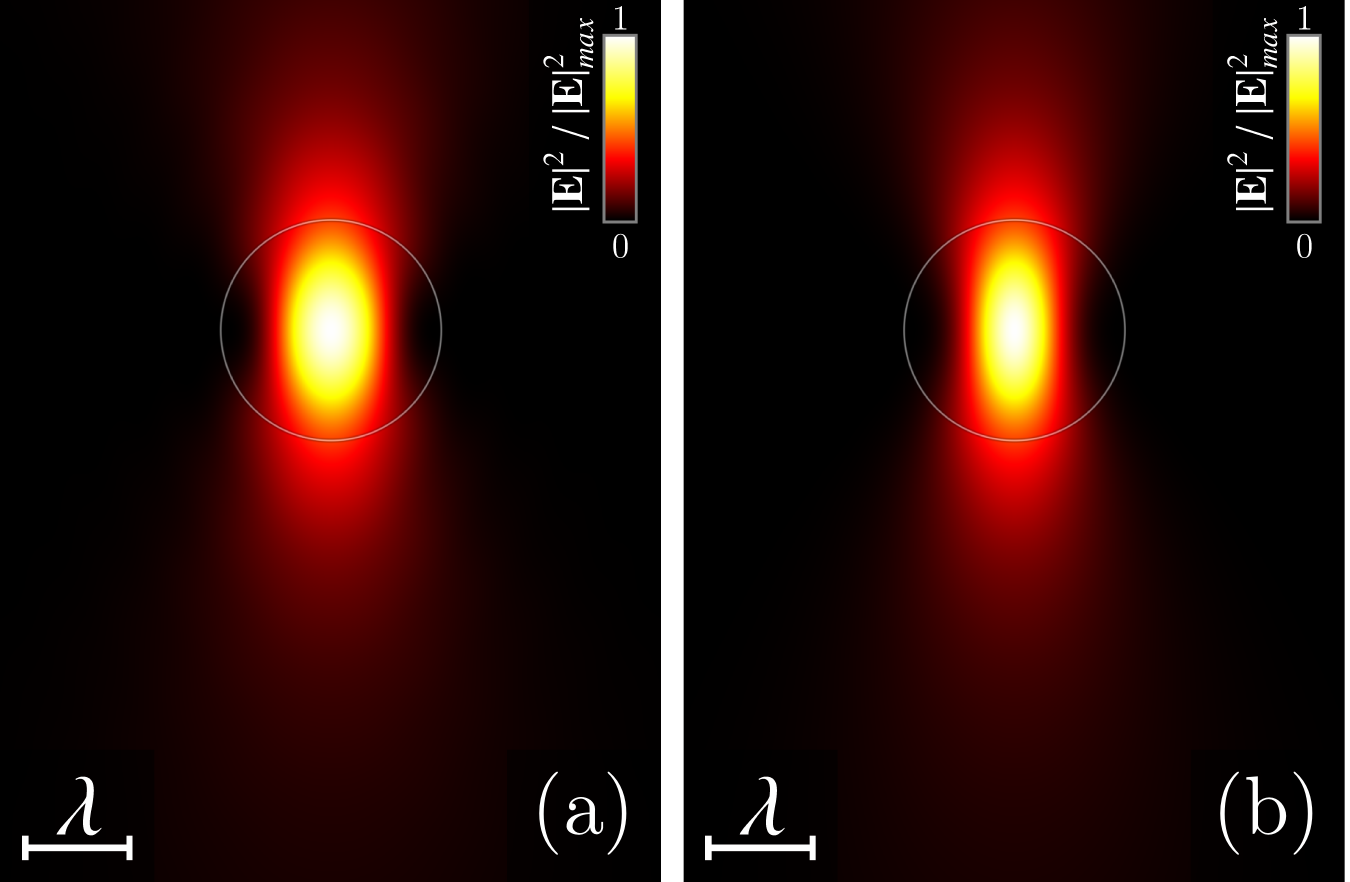}
\caption{%
(Color online)
Intensity distributions of
the incident wave field
in (a)~the $x-z$ plane and
(b)~the $y-z$ plane
for the LG$_{00}$ (Gaussian) beam
with 
$n_{\ind{LG}}=m_{\ind{LG}}=0$ and $f=0.3$.
The $z$ axis is directed from the top down.
}
\label{fig:intensity-00}
\end{figure*}

\begin{figure*}[!tbh]
\includegraphics[width=170mm]{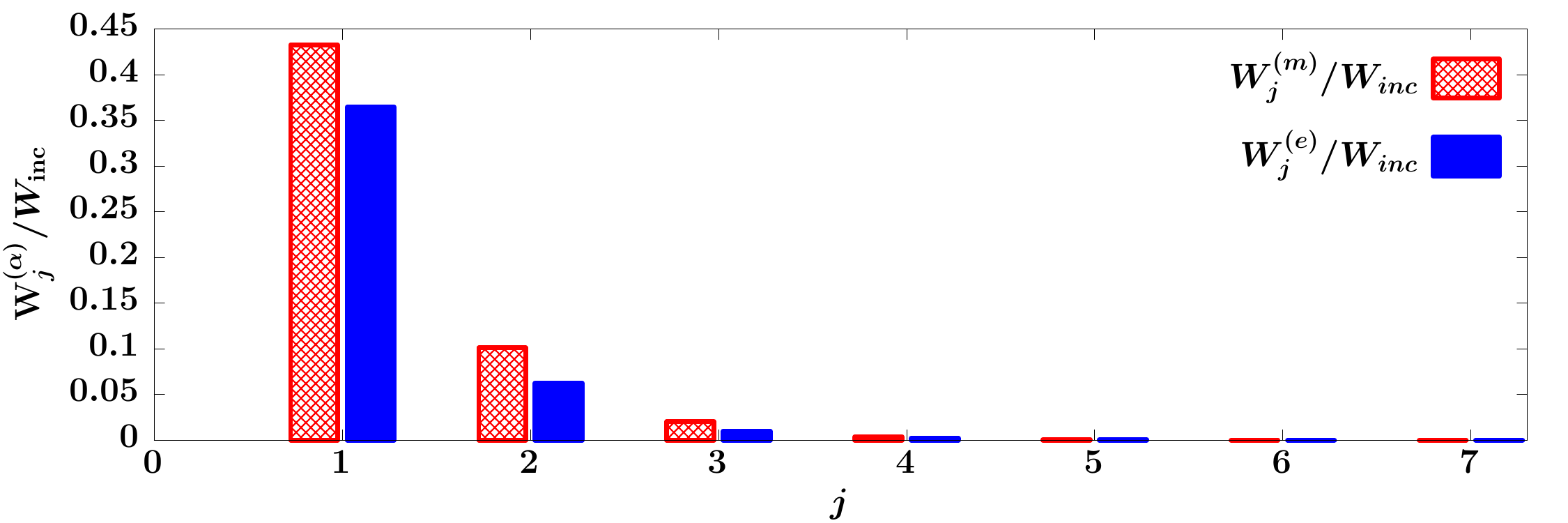}
\caption{%
(Color online)
Histogram of
multipolar decomposition of 
the incident LG$_{00}$ (Gaussian)
beam with $f=0.3$.
Height of the bars represents
relative contribution of 
the modes,
$W_j^{(m)}/W_{\ind{inc}}$
and
  $W_j^{(e)}/W_{\ind{inc}}$,
(see Eqs.~\eqref{eq:W_inc} and~\eqref{eq:W_j})
to the total power
of the incident beam 
depending on
the angular momentum
number $j$.
}
\label{fig:modes-00}
\end{figure*}

\begin{figure*}[!tbh]
\includegraphics[width=170mm]{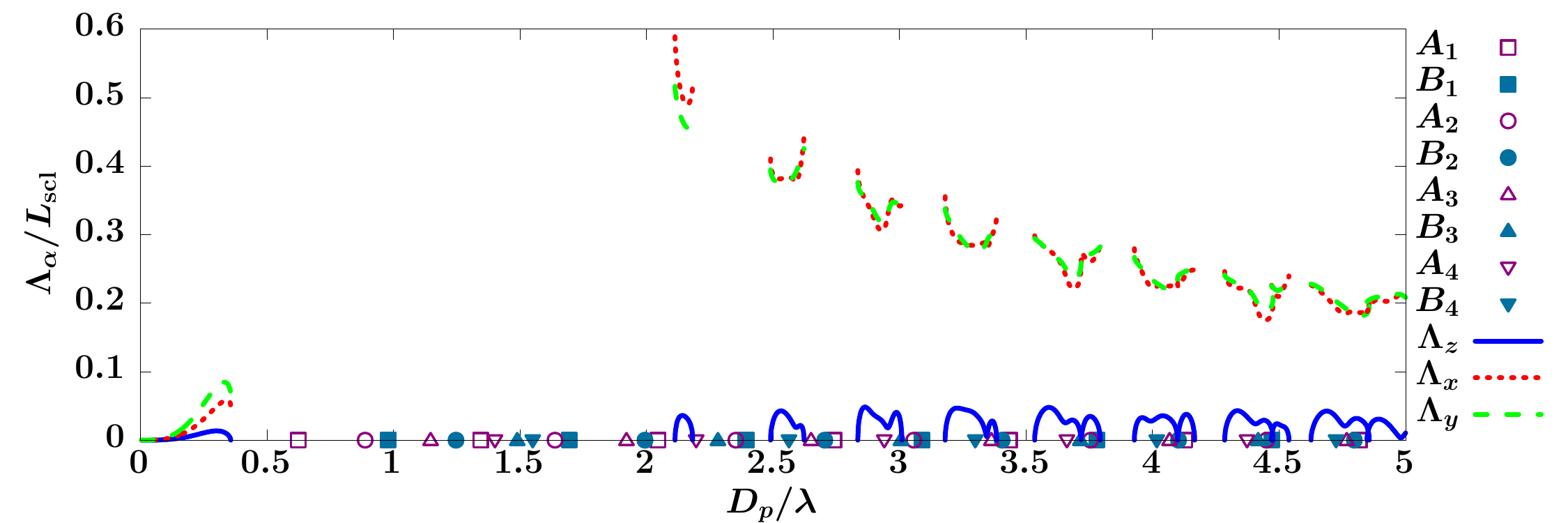}
\caption{%
(Color online)
Eigenvalues of the effective stiffness matrix $\mvc{L}_{\ind{eff}}$
(see Eqs.~\eqref{eq:L-eff} and~\eqref{eq:L-m0}),
$\Lambda_{\alpha}/L_{\ind{scl}}$,
as a function of the size parameter,
$D_p/\lambda=2R_p/\lambda$,
for the LG$_{00}$ beam
with $f=0.3$.
The scale factor is $L_{\ind{scl}}=n W_{\ind{inc}}/(c \lambda m_p)$
(see Eq.~\eqref{eq:L-eff})
and the refractive index of the particle is $n_p=1.33$.
}
\label{fig:eigval-00}
\end{figure*}

\begin{figure*}[!tbh]
\includegraphics[width=170mm]{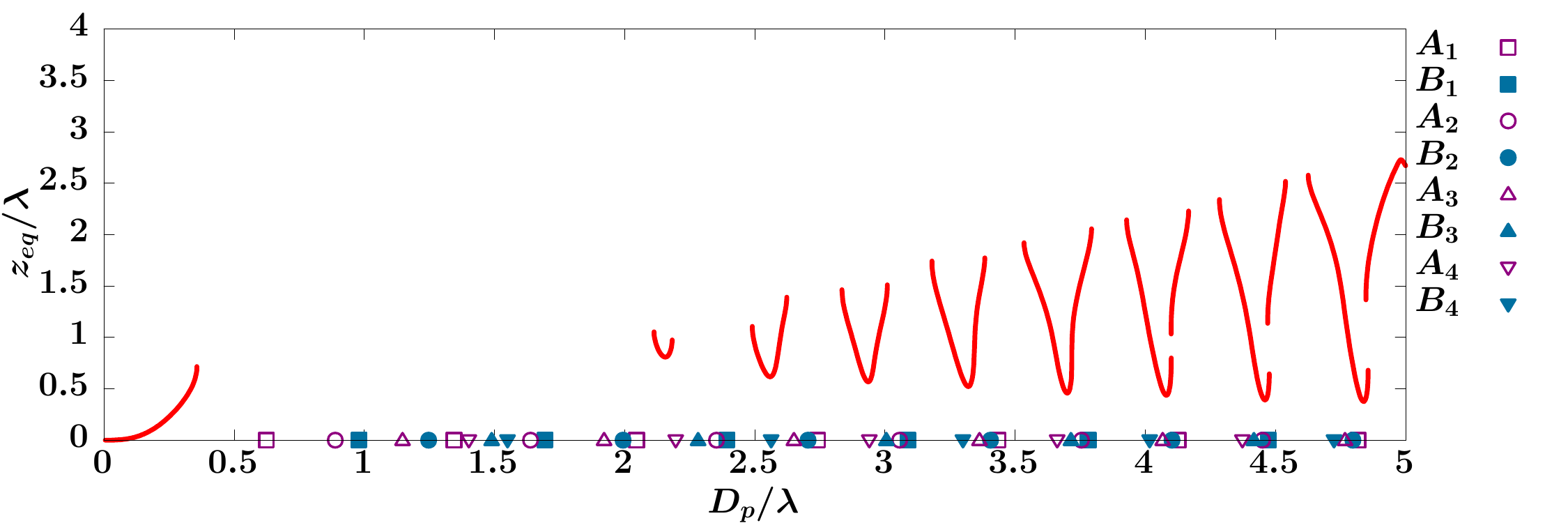}
\caption{%
(Color online)
On-axis coordinate of axially stable zero-force points
$z_{\ind{eq}}$ as a function of
the size parameter for the Gaussian beam.
}
\label{fig:zeq-00}
\end{figure*}

\begin{figure*}[!tbh]
\includegraphics[width=170mm]{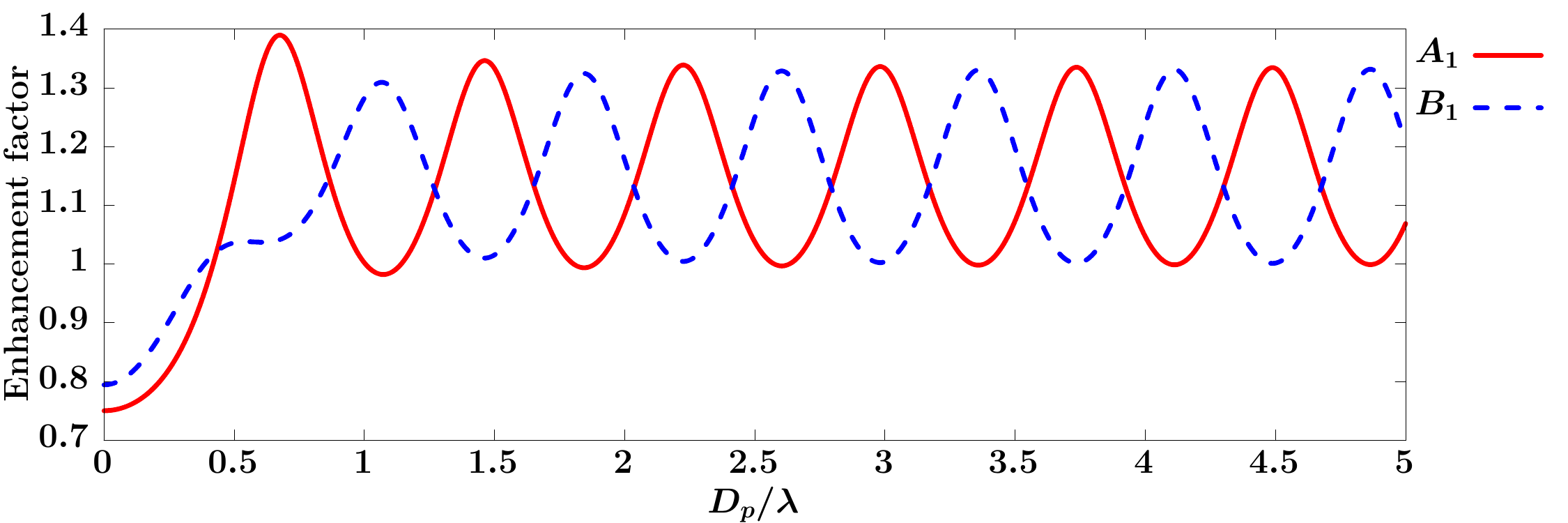}
\caption{%
(Color online)
Enhancement factors,
$A_j=|a_j^{(p)}|^2$
and $B_j=|b_j^{(p)}|^2$,
expressed in terms of
the internal field coefficients
(see Eqs.~\eqref{eq:mie-alp-p} and~\eqref{eq:mie-bet-p})
as a function of the size parameter at $j=1$.
}
\label{fig:mie-1}
\end{figure*}

We consider 
the case where the thermal noise can be neglected
and dynamics of the particle is
governed by the equation of motion
\begin{align}
&
\label{eq:newton}
  \frac{\dd^2 \vc{r}_p}{\dd t^2}+2 \gamma\frac{\dd \vc{r}_p}{\dd
    t}=m_p^{-1} \vc{F}(\vc{r}_p),
\end{align}
where 
$\vc{F}(\vc{r}_p)$ is the optical force
given in Eq.~\eqref{eq:F_far-field};
$\gamma$ is the damping constant
of the ambient medium 
and 
$m_p$ is the mass of the particle.

When the particle is trapped,
it is localized in the vicinity of
a stable equilibrium (steady state)
$\vc{r}_{\ind{eq}}$,
which is the zero-force position where
$\vc{F}(\vc{r}_{\ind{eq}})=\vc{0}$.
Stability of the equilibrium
can be studied in the linear approximation
where Eq.~\eqref{eq:newton}
is approximated by 
the first-order (linearized) dynamic
equations
\begin{align}
&
\label{eq:newton-lin}
  \frac{\dd \vc{v}}{\dd t}+2\gamma
\vc{v}+\mvc{L}_0{\vc{x}}=\vc{0},
\quad
\frac{\dd \vc{x}}{\dd t}=\vc{v},
\\
&
\label{eq:L}
\quad
\mvc{L}_0\equiv
-m_p^{-1} \mvc{K}_{\ind{eq}}
\end{align}
where $\vc{x}=\vc{r}_p-\vc{r}_{\ind{eq}}$
is the displacement vector
and $\mvc{K}_{\ind{eq}}=\mvc{K}(\vc{r}_{\ind{eq}})$
is the stiffness matrix given in Eq.~\eqref{eq:K-def}.

General solution of the system~\eqref{eq:newton-lin}
written in the form 
\begin{align}
&
  \label{eq:Cauchy}
  \begin{pmatrix}
    \vc{x}(t)\\
\vc{v}(t)
  \end{pmatrix}
=
\mvc{U}(t-t_0)
\begin{pmatrix}
    \vc{x}(t_0)\\
 \vc{v}(t_0)
  \end{pmatrix}
\end{align}
describes how the position and the velocity of the particle
evolve in time using
the evolution operator $\mvc{U}(t)$ given by
\begin{align}
&
\label{eq:evol-lin}
\mvc{U}(t)=\ee^{-\gamma t}
\begin{pmatrix}
  \cos\sqrt{\mvc{L}} t & \sqrt{\mvc{L}^{-1}}\sin\sqrt{\mvc{L}}t\\
-\sqrt{\mvc{L}}\sin\sqrt{\mvc{L}}t &  \cos\sqrt{\mvc{L}}t 
\end{pmatrix},
\quad
\mvc{L}=\mvc{L}_0-\gamma^2\mvc{I}_3,  
\end{align}
where $\mvc{I}_3$ is the $3\times 3$
identity matrix. 

If  the evolution operator~\eqref{eq:evol-lin}
contains terms 
that are unbounded functions
of time for $t\in [0,\infty)$,
the equilibrium $\vc{r}_{\ind{eq}}$
is unstable~\cite{Gucken:bk:1990} and 
the particle cannot be trapped
at such a fixed point.
Stability of the equilibrium 
thus requires 
the norm of the matrix exponentials
$\exp[-\gamma\mvc{I}_3 \pm i \sqrt{\mvc{L}}] t$
to be a bounded function of time
and is determined by the spectrum
of the matrix $\mvc{L}$.
More specifically,
for the zero-force point to be stable, 
the eigenvalues of the matrix $\mvc{L}_0$
must satisfy the inequality
\begin{align}
  \label{eq:stability-cond}
  |\Im(\sqrt{\Lambda_i-\gamma^2})|\le\gamma,
\end{align}
where $\Lambda_i$ is the eigenvalue of the matrix
$\mvc{L}_0$.
After some rather straightforward algebraic manipulations,
we can conveniently render the stability
condition~\eqref{eq:stability-cond-gamma}
into the form of the constraint
\begin{align}
  \label{eq:stability-cond-gamma}
  4\gamma^2\Re\Lambda_i\ge[\Im\Lambda_i]^2
\end{align}
imposed on the value of the damping constant
$\gamma$.

Inequality~\eqref{eq:stability-cond-gamma}
suggests that the eigenvalues may generally
be divided into the three groups:
\renewcommand{\theenumi}{\alph{enumi}}
\renewcommand{\labelenumi}{(\theenumi)}
\begin{enumerate}
\item 
at $\Re\Lambda_i<0$,
the point is unstable
and cannot be stabilized
by introducing energy losses
caused by the ambient medium;
\item 
at $\Re\Lambda_i>0$ and $\Im\Lambda_i=0$,
the point is stable even if $\gamma=0$
(the case of vacuum);
\item 
at $\Re\Lambda_i>0$ and $\Im\Lambda_i\ne 0$,
the point is conditionally stable (stabilizable)
meaning that,
even though the point is unstable at $\gamma=0$,
it can be stabilized provided
the particle is embedded into the medium
with sufficiently large damping constant $\gamma$.
\end{enumerate}

Note that an eigenvalue of $\mvc{L}_0$ with
$\Re\Lambda_i=0$ may present 
different cases depending on its imaginary part.
More precisely, the point being conditionally stable
 at $\Im\Lambda_i=0$ would be unstable otherwise.

Another remark concerns the non-generic case 
when the matrix $\mvc{L}_0$ is not diagonalizable
and its Jordan normal form 
contains a Jordan block.
This may happen only if there are repeated eigenvalues
of $\mvc{L}_0$
which geometric multiplicity
is strictly less then the algebraic one.
As opposed to the case of diagonalizable matrix,
at the boundary of the stability region where
$|\Im(\sqrt{\Lambda_i-\gamma^2})|=\gamma$,
the exponentials $\exp[-\gamma\mvc{I}_3 \pm i \sqrt{\mvc{L}}] t$
will diverge at $t\to\infty$ and 
the zero-force point is unstable.

Now, similar to the force efficiency~\eqref{eq:force-efficiency},
we introduce the dimensionless effective stiffness matrix
\begin{align}
  \label{eq:L-eff}
  \mvc{L}_{\ind{eff}}=\mvc{L}_0/L_{\ind{scl}},
\quad
L_{\ind{scl}}=F_{\ind{scl}}/(\lambda m_p),
\end{align}
where the force scale factor $F_{\ind{scl}}$
is given in Eq.~\eqref{eq:force-efficiency},
and present the results of our numerical analysis
for the technologically important case
of fixed points located on the laser beam axis (the $z$ axis),
$\vc{r}_{\ind{eq}}=(0,0,z_{\ind{eq}})$.

\begin{figure*}[!tbh]
\includegraphics[width=120mm]{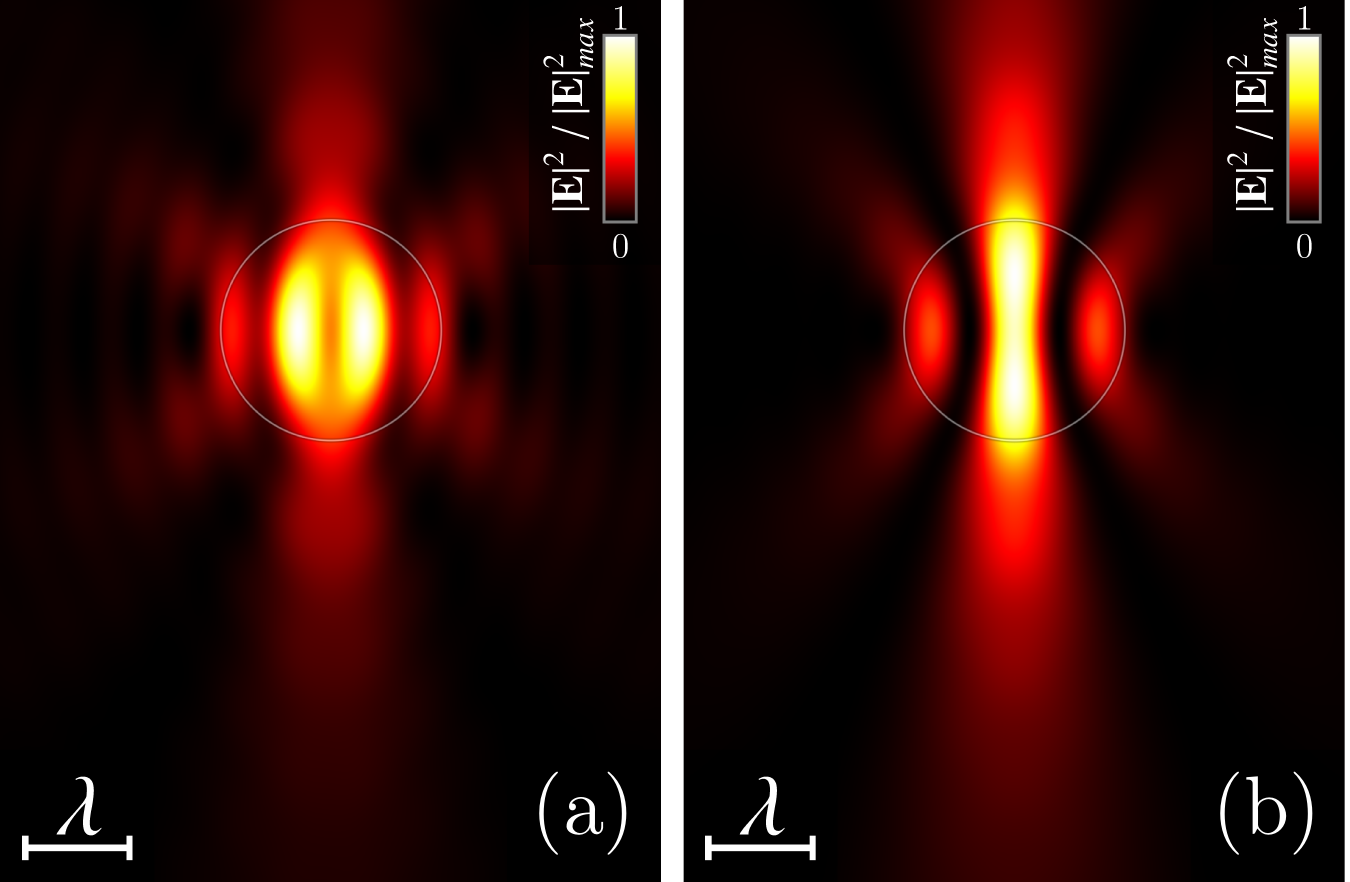}
\caption{%
(Color online)
Intensity distributions of
the incident wave field
in (a)~the $x-z$ plane and
(b)~the $y-z$ plane
for the non-vertex LG$_{10}$ beam
with $f=0.3$.
The $z$ axis is directed from the top down.
}
\label{fig:intensity-10}
\end{figure*}

\begin{figure*}[!tbh]
\includegraphics[width=170mm]{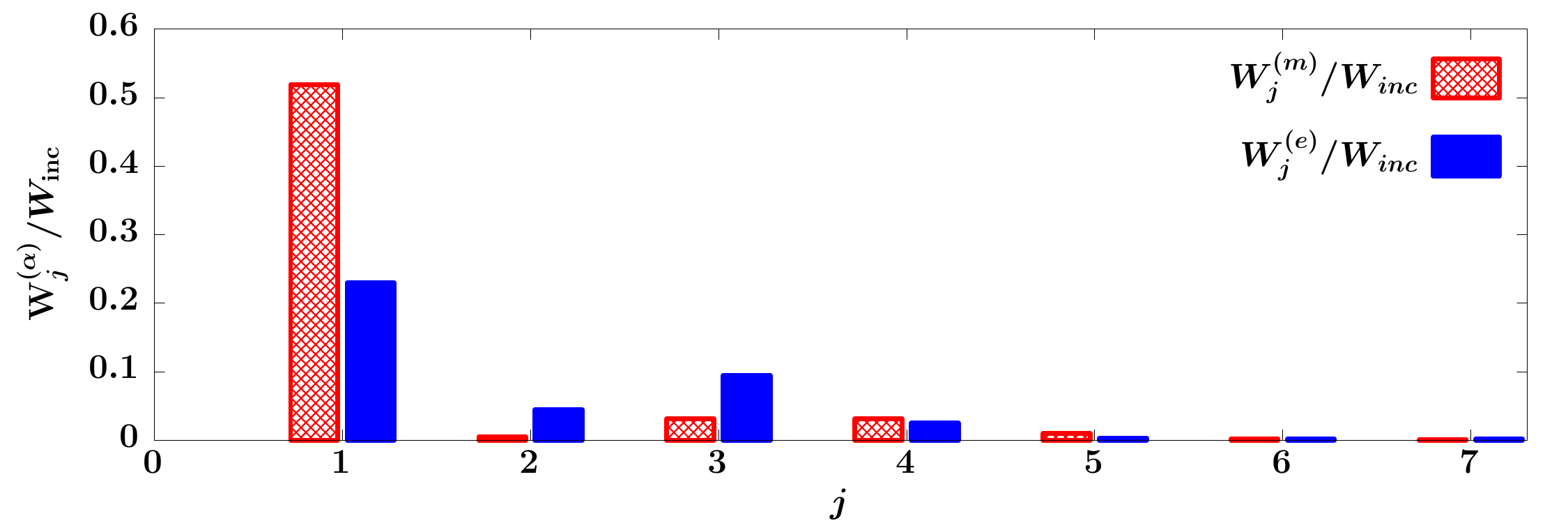}
\caption{%
(Color online)
Histogram of
multipolar decomposition of 
the incident LG$_{10}$
beam with $f=0.3$.
Height of the bars represents
relative contribution of 
the modes,
$W_j^{(m)}/W_{\ind{inc}}$
and
  $W_j^{(e)}/W_{\ind{inc}}$,
(see Eqs.~\eqref{eq:W_inc} and~\eqref{eq:W_j})
to the total power
of the incident beam 
depending on
the angular momentum
number $j$.
}
\label{fig:modes-10}
\end{figure*}

\begin{figure*}[!tbh]
\includegraphics[width=170mm]{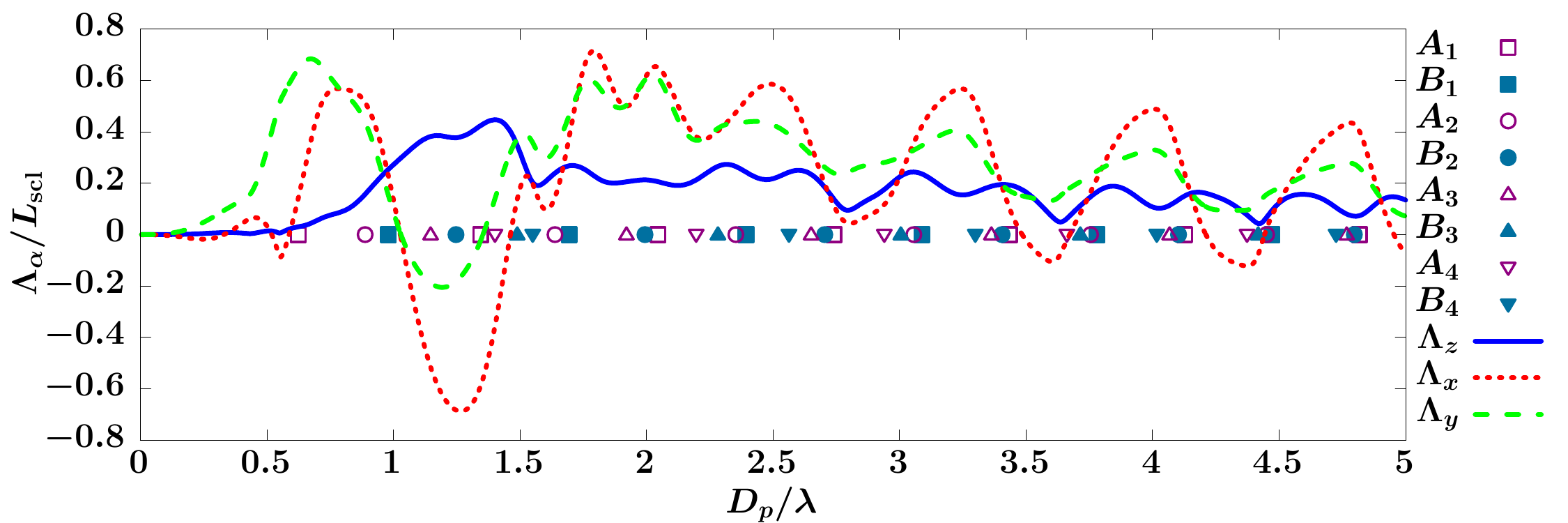}
\caption{%
(Color online)
Eigenvalues of the effective stiffness matrix $\mvc{L}_{\ind{eff}}$
(see Eqs.~\eqref{eq:L-eff} and~\eqref{eq:L-m0}),
$\Lambda_{\alpha}/L_{\ind{scl}}$,
as a function of the size parameter
for the LG$_{10}$ beam
with $f=0.3$.
The scale factor is $L_{\ind{scl}}=n W_{\ind{inc}}/(c \lambda m_p)$
(see Eq.~\eqref{eq:L-eff}).
}
\label{fig:eigval-10}
\end{figure*}

\begin{figure*}[!tbh]
\includegraphics[width=170mm]{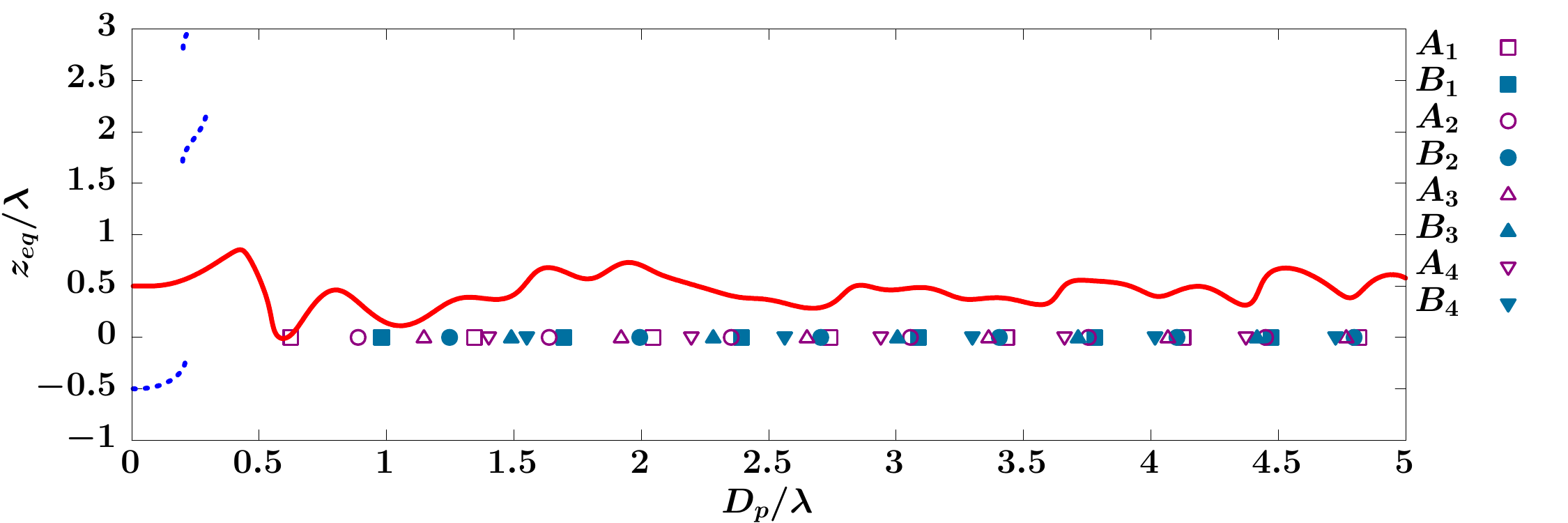}
\caption{%
(Color online)
On-axis coordinate of axially stable zero-force points
$z_{\ind{eq}}$ as a function of
the size parameter for the LG$_{10}$ beam. 
}
\label{fig:zeq-10}
\end{figure*}

\subsection{Non-vortex beams with $m_{\ind{LG}}=0$}
\label{subsec:non-vortex}

We begin with the results for non-vortex LG beams
characterized by the vanishing azimuthal mode
number $m_{\ind{LG}}=0$. 
The well known example of such beams is the
Gaussian beam, LG$_{00}$,
where the radial mode number is also equal to zero.
Figure~\ref{fig:intensity-00} shows the
two-dimensional
(2D) intensity distributions in the $x-z$ and $y-z$ plane
for the LG$_{00}$ beam with the focusing parameter
$f=0.3$. 
Multipolar decomposition representing
the total power of the incident LG$_{00}$ beam
resolved into the contributions from the electric and magnetic
modes with different angular momentum number $j$
(see Eq.~\eqref{eq:W_inc}) is presented in
Fig.~\ref{fig:modes-00}. 

From our symmetry analysis
performed in Sec.~\ref{subsec:symmetry},
for the non-vortex beams, the stiffness
matrix is diagonal (see Eq.~\eqref{eq:K-m0}).
So, the matrix~\eqref{eq:L} takes the diagonal form:
\begin{align}
  \label{eq:L-m0}
  \mvc{L}_0=
-m_p^{-1}
\diag(K_{xx}^{(\ind{eq})},  K_{yy}^{(\ind{eq})},  K_{zz}^{(\ind{eq})}) 
\equiv
\diag(\Lambda_{x},\Lambda_{y},\Lambda_{z}),
\end{align}
where the eigenvalues 
are equal to the real-valued diagonal elements
of $\mvc{L}_0$.

In the linear approximation,
these eigenvalues dictate the dynamical
regime of the particle motion 
along the coordinate axes.
In particular, the longitudinal mode
governed by the eigenvalue 
\begin{align}
  \label{eq:lambda-z}
\Lambda_z=-m_p^{-1}K_{zz}^{(\ind{eq})} 
\end{align}
determine the axial stability of 
the zero-force point.
In what follows we confine our analysis to 
the case of the axially stable equilibrium points
with $\Lambda_z\ge 0$.
The results for these points
are shown in Figs.~\ref{fig:eigval-00}
and~\ref{fig:zeq-00}.
Referring to Fig.~\ref{fig:eigval-00},
the transverse eigenvalues $\Lambda_{x}$ and $\Lambda_{y}$
being close to each other
are considerably greater than the longitudinal one:  
$\Lambda_{x}\approx\Lambda_{y}>\Lambda_{z}$.
So, it turned out that all the axially stable equilibria are 
the trapping points (stable zero-force points).
The coordinate of the trapping point plotted in relation to the
size parameter of the particle, 
$D_p/\lambda=2 R_p/\lambda$,
is depicted in Fig.~\ref{fig:zeq-00}.

In Figs.~\ref{fig:eigval-00}
and~\ref{fig:zeq-00},
differently shaped marks 
are used to indicate the Mie resonance
values of the scatterer size parameter
for various modes.
Such resonances also known as the
morphology-dependent resonances
(the whispering gallery modes)
reveal themselves in non-monotonic
oscillating behavior of 
the magnitude of
the internal field coefficients
given by Eqs.~\eqref{eq:mie-alp-p} and~\eqref{eq:mie-bet-p}.
For 
the enhancement factors defined as the square
of the modulus of
the internal field coefficients, 
$A_j=|a_j^{(p)}|^2$
and $B_j=|b_j^{(p)}|^2$,
with
$j=1$,
such oscillations can be seen 
in Fig.~\ref{fig:mie-1}.
Open and filled squares
are used to mark the values
of the size parameter $D_p/\lambda$
corresponding to local maxima of 
the enhancement factors $A_1$ and $B_1$,
respectively.

The LG$_{10}$ beam
characterized by the intensity distributions
and the multipolar decomposition
shown in Figs.~\ref{fig:intensity-10} and~\ref{fig:modes-10},
respectively,
presents the case of a non-vertex 
incident beam with nonzero radial mode number.
By contrast to the case of the Gaussian beams,
as is seen from Fig.~\ref{fig:eigval-10},
the longitudinal eigenvalue, $\Lambda_z$,
and the transverse stiffness coefficients,
$\Lambda_x$ and $\Lambda_y$,
are of the same order.

Referring to Fig.~\ref{fig:eigval-10},
for the LG$_{10}$ beam,
stability of equilibria is determined 
by the sign of the transverse eigenvalue $\Lambda_x$,
whereas, for the Gaussian beam,
the stability governing factor is the sign of $\Lambda_z$.
In addition, 
the size parameter dependence of 
the zero-force point coordinate shown in
Fig.~\ref{fig:zeq-10} demonstrates the presence
of several branches of axially stable equilibria
in the region of subwavelength scatterers.

\begin{figure*}[!tbh]
\includegraphics[width=120mm]{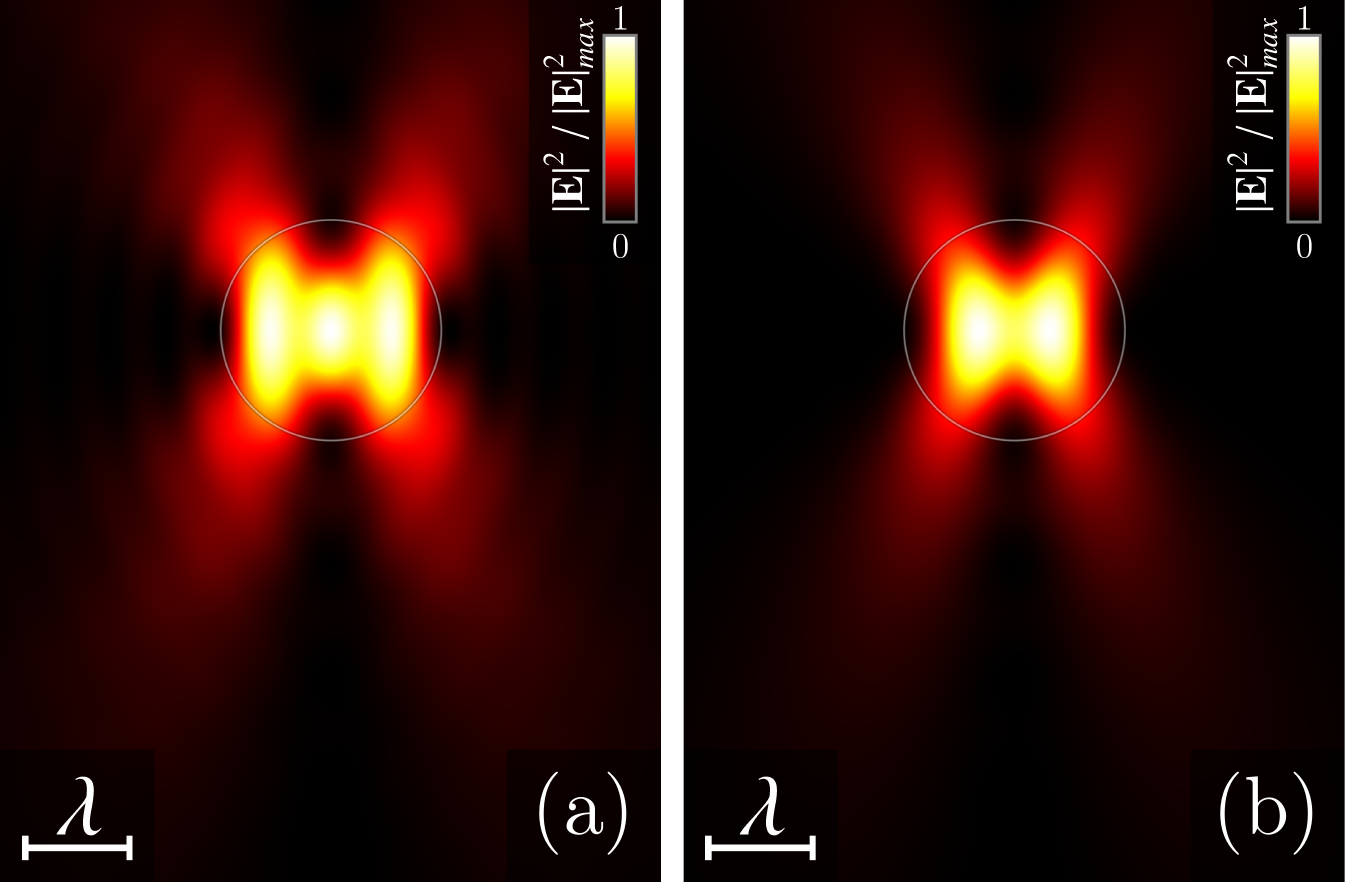}
\caption{%
(Color online)
Intensity distributions of
the incident wave field
in (a)~the $x-z$ plane and
(b)~the $y-z$ plane
for the LG$_{01}$ beam
with $f=0.3$.
The $z$ axis is directed from the top down.
}
\label{fig:intensity-01}
\end{figure*}

\begin{figure*}[!tbh]
\includegraphics[width=170mm]{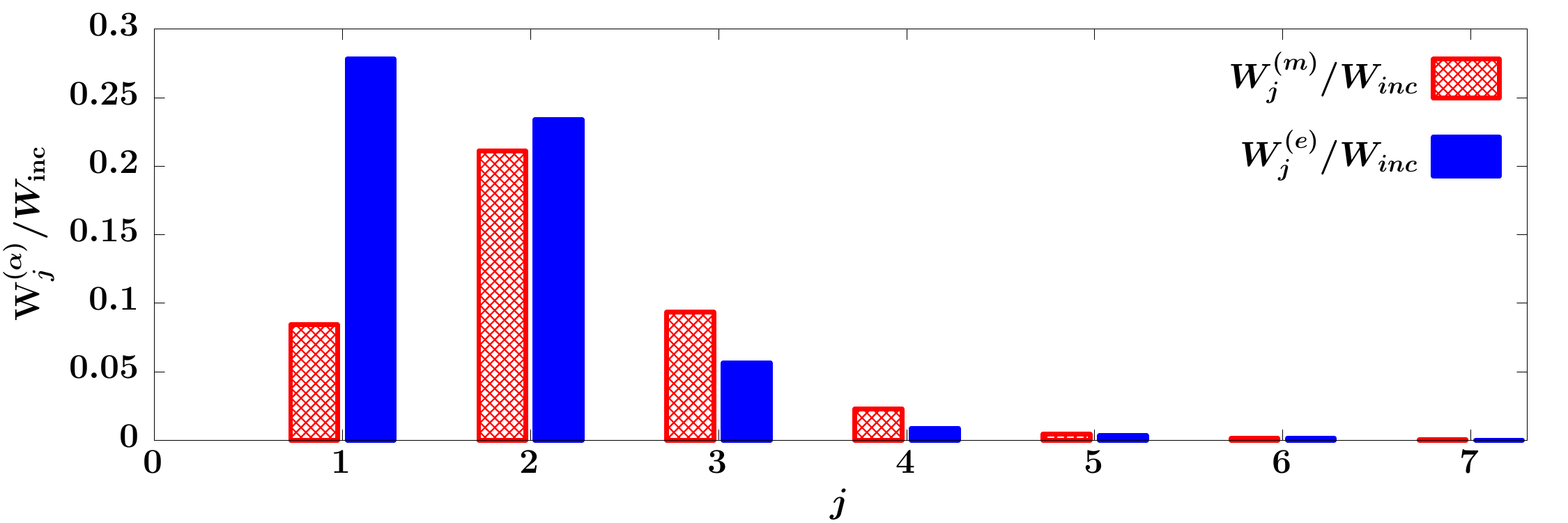}
\caption{%
(Color online)
Histogram of
multipolar decomposition of 
the incident LG$_{01}$
beam with $f=0.3$.
Height of the bars represents
relative contribution of 
the modes,
$W_j^{(m)}/W_{\ind{inc}}$
and
  $W_j^{(e)}/W_{\ind{inc}}$,
(see Eqs.~\eqref{eq:W_inc} and~\eqref{eq:W_j})
to the total power
of the incident beam 
depending on
the angular momentum
number $j$.
}
\label{fig:modes-01}
\end{figure*}

\begin{figure*}[!tbh]
\includegraphics[width=170mm]{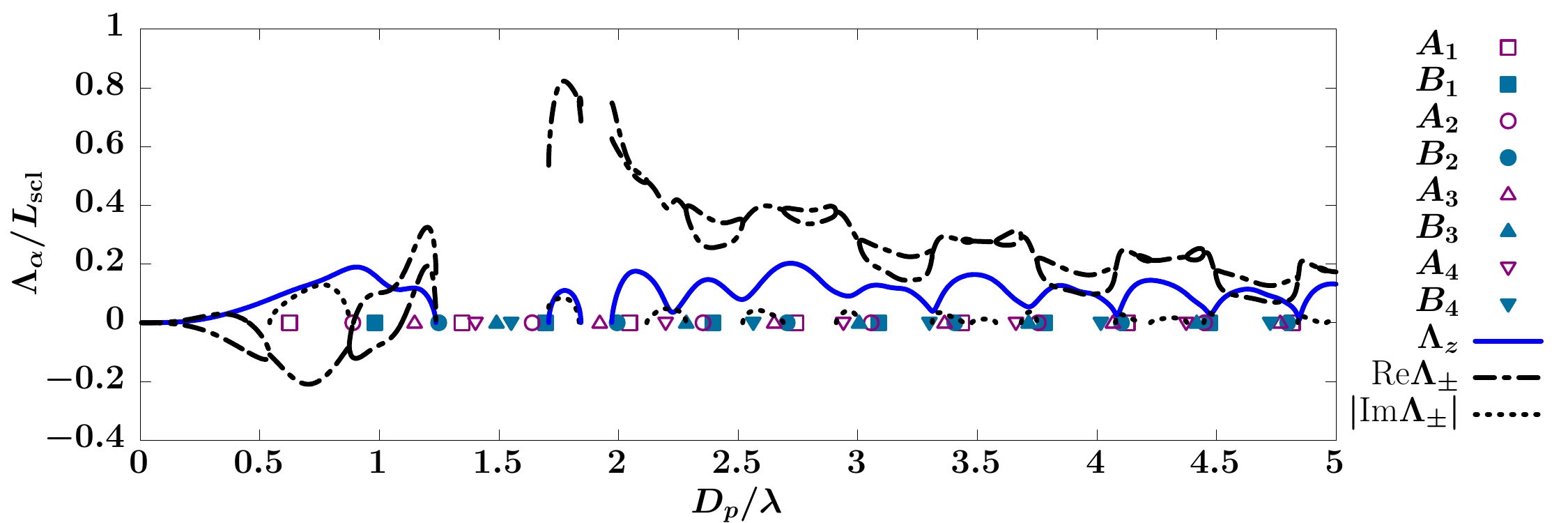}
\caption{%
(Color online)
Eigenvalues of the effective stiffness matrix $\mvc{L}_{\ind{eff}}$
(see Eqs.~\eqref{eq:L-eff} and~\eqref{eq:L-on-axis}),
$\Lambda_{\alpha}/L_{\ind{scl}}$,
as a function of the size parameter
for the LG$_{01}$ beam
with $f=0.3$.
}
\label{fig:eigval-01}
\end{figure*}

\begin{figure*}[!tbh]
\includegraphics[width=170mm]{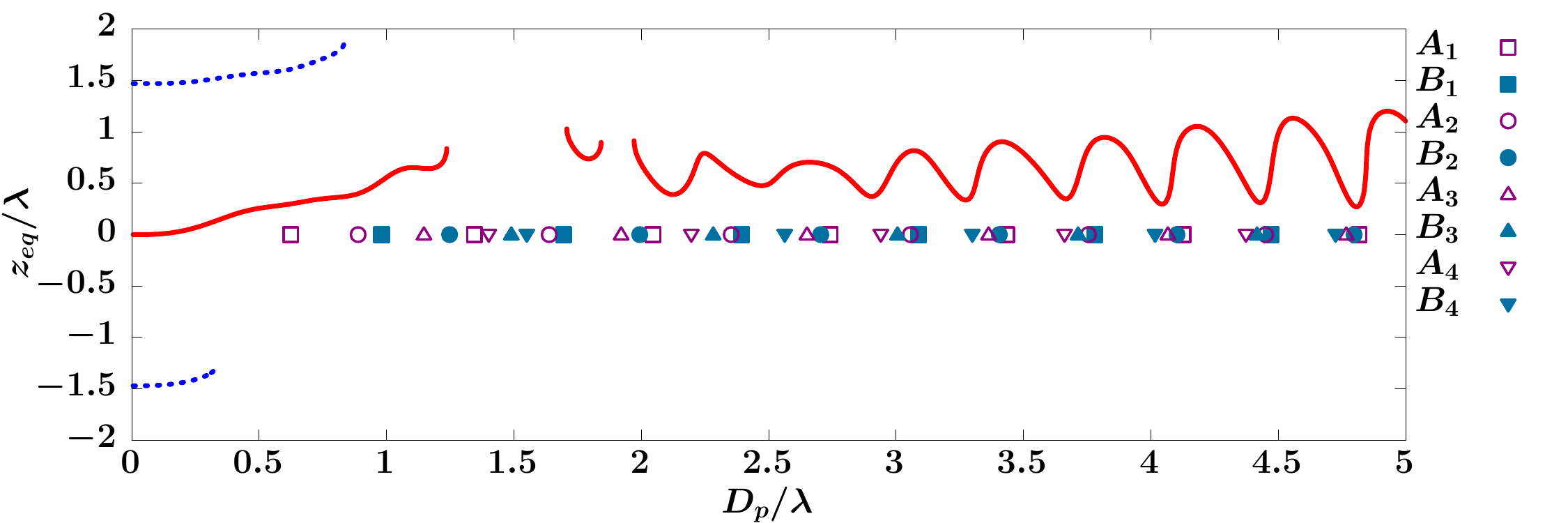}
\caption{%
(Color online)
On-axis coordinate of axially stable zero-force points
$z_{\ind{eq}}$ as a function of
the size parameter for the LG$_{01}$ beam. 
}
\label{fig:zeq-01}
\end{figure*}

\begin{figure*}[!tbh]
\includegraphics[width=120mm]{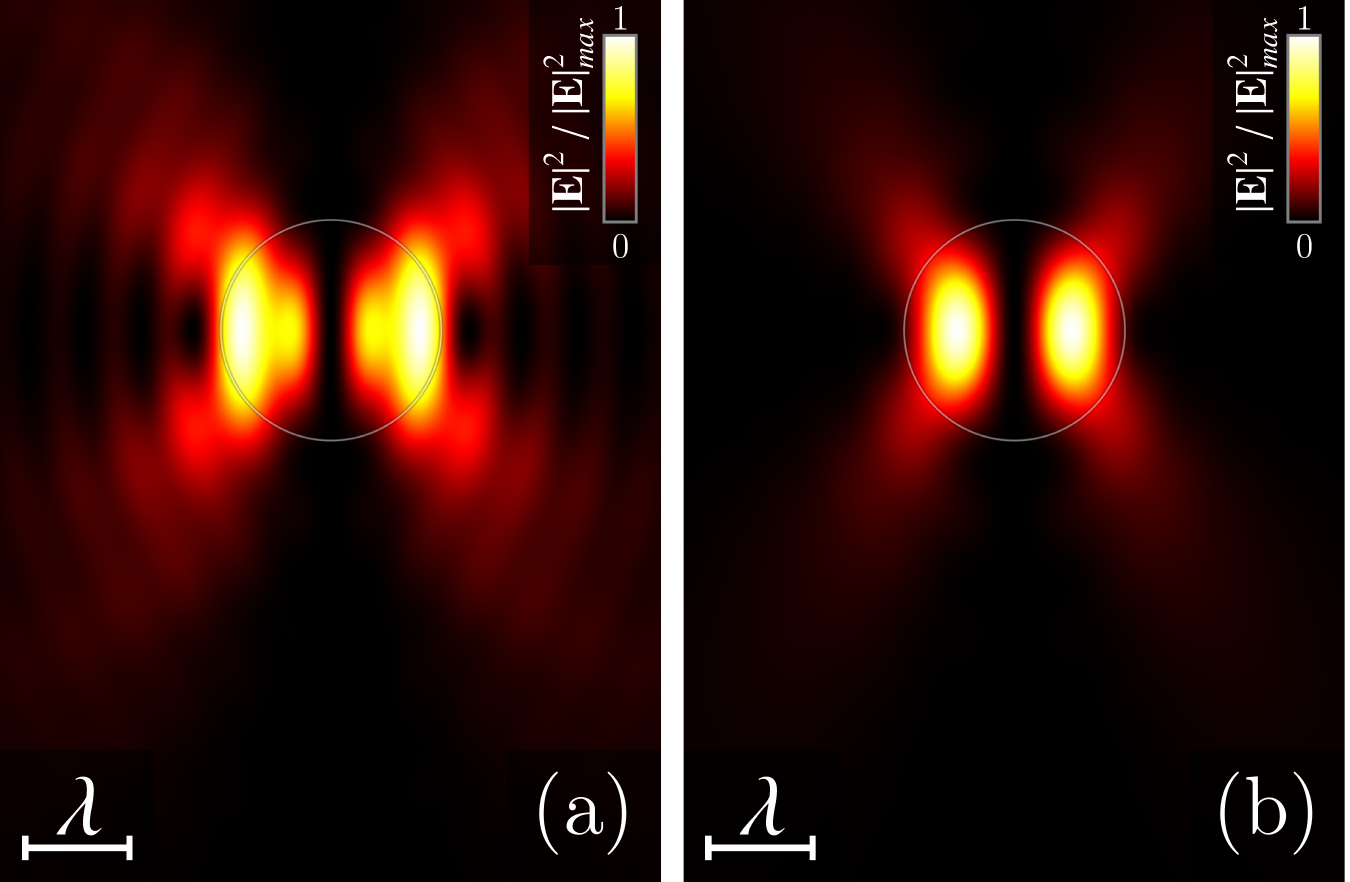}
\caption{%
(Color online)
Intensity distributions of
the incident wave field
in (a)~the $x-z$ plane and
(b)~the $y-z$ plane
for the LG$_{02}$ beam
with $f=0.3$.
The $z$ axis is directed from the top down.
}
\label{fig:intensity-02}
\end{figure*}

\begin{figure*}[!tbh]
\includegraphics[width=170mm]{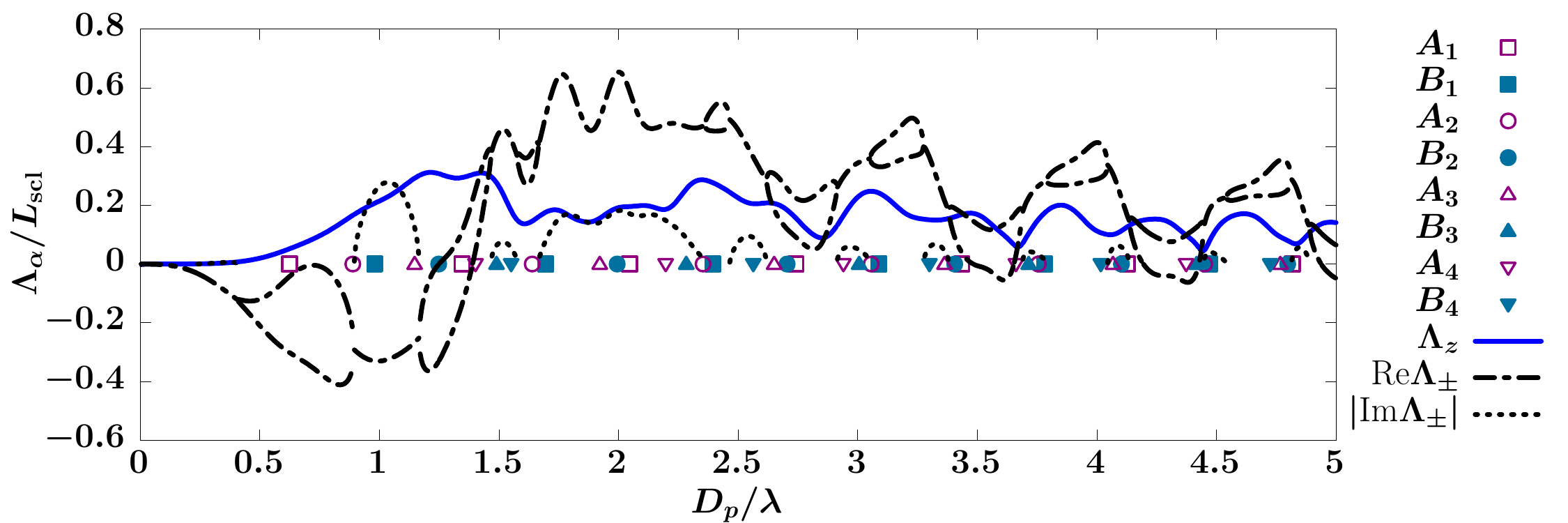}
\caption{%
(Color online)
Eigenvalues of the effective stiffness matrix $\mvc{L}_{\ind{eff}}$
(see Eqs.~\eqref{eq:L-eff} and~\eqref{eq:L-on-axis}),
$\Lambda_{\alpha}/L_{\ind{scl}}$,
as a function of the size parameter
for the LG$_{02}$ beam
with $f=0.3$.
}
\label{fig:eigval-02}
\end{figure*}

\begin{figure*}[!tbh]
\includegraphics[width=170mm]{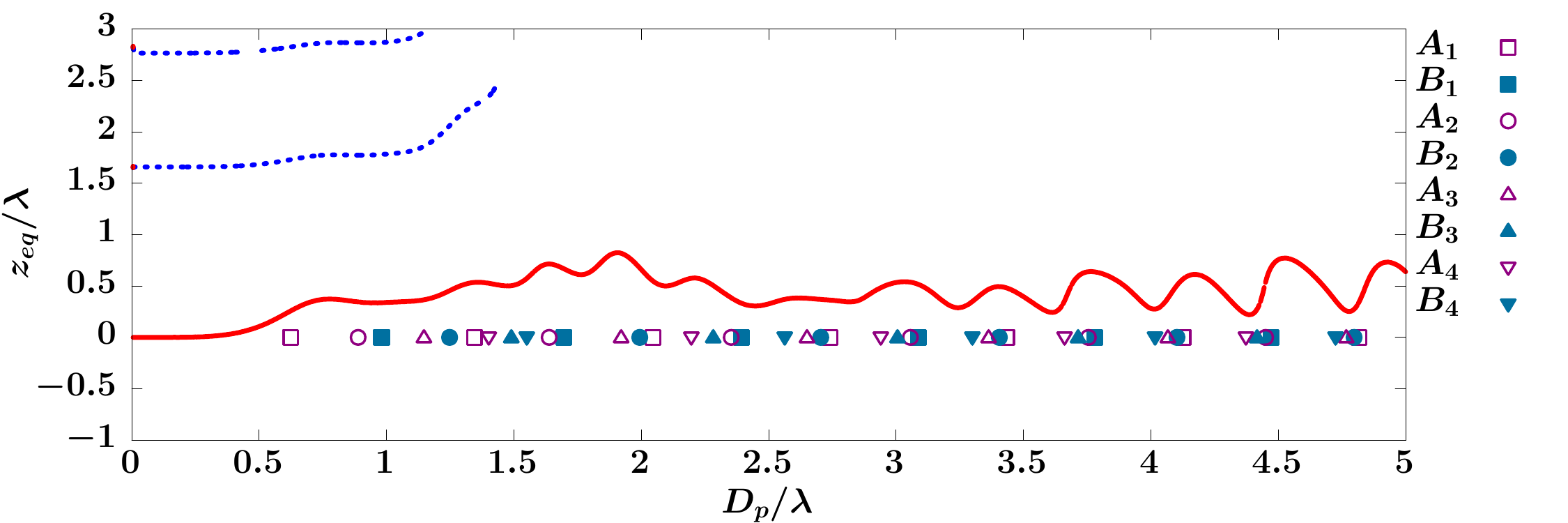}
\caption{%
(Color online)
On-axis coordinate of axially stable zero-force points
$z_{\ind{eq}}$ as a function of
the size parameter for the LG$_{02}$ beam. 
}
\label{fig:zeq-02}
\end{figure*}

\subsection{Optical vortex beams: effects of non-conservative dynamics}
\label{subsec:vortex}

It should be stressed that, for
the above discussed case of non-vortex beams
is characterized by
the symmetric stiffness matrix
and the dynamics of the particle 
is thus locally conservative.

Since all the eigenvalues of such matrices are real,
there are no conditionally stable equilibria and 
stability of all the zero-force points turned out to
be essentially independent of the ambient damping.
For the laser beams carrying a phase singularity
known as the vortex 
the latter is no longer the case.

The topological charge characterizing the phase singularity
and associated orbital angular momentum
are known to
produce a rich variety of phenomena~\cite{Andrews:bk:2008}
such as
rotation of trapped spheres by vortex
beams~\cite{Grier:nat:2003,Simpson:josaa:1:2009}. 
The latter 
is a remarkable manifestation of
the non-conservative nature of optical-force-induced dynamics
meaning that 
optical forces cannot generally be derived from an underlying
potential. The optical force field includes a
scattering contribution, and asymmetric couplings will occur
between coordinates which will lead to asymmetric stiffness
matrices~\cite{Simpson:pre:2010,Ng:prl:2010}.

In this section, 
we consider 
purely azimuthal LG beams~\cite{Rury:pra:2012}
with $n_{\ind{LG}}=0$
and $m_{\ind{LG}}\ne 0$
that represent optical vortex beams.
Symmetry of such beams
has been discussed in Sec.~\ref{subsec:symmetry}
leading to the conclusion that
the linearized dynamics is governed 
by the non-symmetric stiffness matrix of the form:
\begin{align}
  \label{eq:L-on-axis}
  \mvc{L}_0=
-m_p^{-1}
 \begin{pmatrix}
    K_{xx}^{(\ind{eq})} & K_{xy}^{(\ind{eq})} & 0\\
K_{yx}^{(\ind{eq})} & K_{yy}^{(\ind{eq})} & 0\\
0 & 0 & K_{zz}^{(\ind{eq})}
  \end{pmatrix}
=
\diag(\mvc{L}_t,\Lambda_z),
\\
  \label{eq:Lt-on-axis}
\quad
\mvc{L}_t=
\begin{pmatrix}
  L_{11}&L_{12}\\
L_{21}&L_{22}
\end{pmatrix}
=-m_p^{-1}
 \begin{pmatrix}
    K_{xx}^{(\ind{eq})} & K_{xy}^{(\ind{eq})}\\
K_{yx}^{(\ind{eq})} & K_{yy}^{(\ind{eq})}\\
  \end{pmatrix}.
\end{align}
Formula~\eqref{eq:L-on-axis} shows
that, similar to the case of non-vortex beams,
the eigenvalue $\Lambda_z$ given by Eq.~\eqref{eq:lambda-z}
controls axial stability of the equilibria
whereas  
the eigenvalues of the matrix~\eqref{eq:Lt-on-axis}
(the transverse eigenvalues)
\begin{align}
  \label{eq:lambda-pm}
  \Lambda_{\pm}=L_{+}\pm\sqrt{L_{-}^2+L_{12}L_{21}},
\quad
L_{\pm}=(L_{11}\pm L_{22})/2
\end{align}
dictate the dynamics in the transverse plane
(the $x-y$ plane) and govern the radial (transverse)
stability of the zero-force points.

Figures~\ref{fig:intensity-01} and~\ref{fig:modes-01}
present the intensity distributions
and the mode decomposition for the 
focused LG$_{01}$ beam remodelled using 
the focusing parameter $f=0.3$.
The eigenvalues and the coordinate of 
the axially stable zero-force point computed as a function of 
the size parameters are shown in
Figs.~\ref{fig:eigval-01} and~\ref{fig:zeq-01},
respectively.

From the plots depicted in Fig.~\ref{fig:eigval-01},
the zero-force point is axially unstable
in the two intervals which upper boundary points
appear to be close to the size ratio
$D_p/\lambda$
corresponding the local maxima
(the Mie resonances)
of the enhancement factors $B_2$ 
($D_p/\lambda\approx 1.25$ and $D_p/\lambda\approx 2.0$)
and $B_1$ ($D_p/\lambda\approx 1.65$). 
In the remaining part of the size parameter region,
stability is determined by
the transverse eigenvalues~\eqref{eq:lambda-pm}.

Referring to Fig.~\ref{fig:eigval-01},
the interval separating the regions of axial instability,
represent the conditionally unstable points
with $\Re\Lambda_{+}=\Re\Lambda_{-}>0$
and $\Im\Lambda_{\pm}\ne 0$.
By contrast,
in the region of small particles,
the equilibrium points are mainly unstable
except for the small interval of stable points
($\Re\Lambda_{+}>\Re\Lambda_{-}>0$ and
$\Im\Lambda_{\pm}=0$)
located 
below the Mie resonance point at $D_p/\lambda\approx 1.25$. 

For larger particles,
at $D_p/\lambda> 1.65$,
the $\Lambda_{\pm}$ curves indicate
the presence of both
stable and conditionally stable trapping points.
This is the region where, 
as it can be seen from Fig.~\ref{fig:zeq-01},
the size dependence
of the equilibrium coordinate $z_{eq}$ shows
increasingly oscillating behavior with
minima located near certain Mie resonance values
of the size parameter.

Note that, in the 2D distributions
for the beams with $m_{\ind{LG}}=1$
shown in Fig.~\ref{fig:intensity-01},
the intensity is clearly
nonzero on the $z$ axis 
in the near-field region localized 
inside the scatterer.
According to Ref.~\cite{Kiselev:pra:2014},
the near-field contributions to the electric field
that are 
linearly polarized along the propagation 
axis are responsible for destroying
the vortex.

As is evident from
the intensity distributions
presented in Fig.~\ref{fig:intensity-02},
there are no such effects for the LG$_{02}$
beams with $m_{\ind{LG}}=2$.  
For such beams, the results for 
the eigenvalues of the stiffness
matrix plotted in Fig.~\ref{fig:eigval-02}
indicate that, similar to the LG$_{10}$ beams
(see Fig.~\ref{fig:eigval-10}), 
all the zero-force points are axially stable.
It can also be seen that
the endpoints of the instability and conditional stability
intervals correlate well with the Mie resonances.
As in the case of the LG$_{01}$ beams,
in the large particle region with $D_p/\lambda>3$,
the size dependence of
the equilibrium coordinate
shown in Fig.~\ref{fig:zeq-02}
reveals the oscillating regime
with minima related to the resonances.
Another effect shared by
all the non-Gaussian beams under consideration
is the presence of additional branches of axially stable equilibria
provided the size of the particle is sufficiently small
(see Figs.~\ref{fig:zeq-10},~\ref{fig:zeq-01} and~\ref{fig:zeq-02}).
These branches, however, predominantly represent 
radially unstable equilibria and
we have omitted 
the corresponding results for the eigenvalues.

\section{Conclusions}
\label{sec:conclusions}

In this paper, 
we have studied the optical-force-induced dynamics
of Mie particles illuminated with LG beams.
For this purpose,
we have used
a $T$--matrix approach in the form
described in Refs.~\cite{Kiselev:pre:2002,Kiselev:pra:2014}.
Our approach also uses  
the remodeling procedure
in which the far-field matching method is
combined with  the results for nonparaxial propagation of
LG beams.
Scattering of such beams is thus described
in terms of the far-field angular distributions,
 $\vc{E}_{\ind{out}}^{\ind{(inc)}}$
and $\vc{E}_{\ind{out}}^{\ind{(sca)}}$,
that determine the outgoing parts of the incident and scattered
waves 
[see Eqs.~\eqref{eq:E_out_expan} and~\eqref{eq:E_out_sca_expan}].
The far-field distributions 
play the central part in the method giving,
in particular, 
the differential cross-sections
[see Eqs.~\eqref{eq:S_tot_flux} and~\eqref{eq:W_sca_ext}]
and the optical (radiation) force acting upon the Mie scatterer
[see Eq.~\eqref{eq:F_far-field}].

The symmetry analysis performed
in Sec.~\ref{subsec:opt-force}
for the LG beams
with the far-field distribution given by
Eq.~\eqref{eq:E_out_LG}
have shown that,
owing to the twofold rotational symmetry
[see Eq.~\eqref{eq:LG-C2}],
the stiffness matrix~\eqref{eq:K-def}
is generally non-symmetric and non-diagonal 
[see Eq.~\eqref{eq:K-C2}]
provided the LG beam
carries the optical vortex
with the topological charge characterized by the azimuthal 
number $m_{\ind{LG}}$.
By contrast, for the non-vortex LG beams
with $m_{\ind{LG}}=0$,
the stiffness matrix is diagonal
(see Eq.~\eqref{eq:K-m0}).
The form of the beam shape
coefficients~\eqref{eq:alpha-beta-inc-rules}
is also dictated by the twofold rotational
symmetry of the LG beam.

The analytical results
for 
the optical force and the stiffness matrix
are employed to
perform numerical analysis of
the dynamics of the particle
embedded into the viscous medium characterized
by the damping constant $\gamma$
[the equation of motion is given by Eq.~\eqref{eq:newton}].
In this analysis, we have examined stability of 
the zero-force axial points
and  the associated regimes of the linearized dynamics
governed by Eq.~\eqref{eq:newton-lin}.
These regimes are shown to be dictated by 
the eigenvalues of the stiffness matrix~\eqref{eq:L}
that enter the stability condition~\eqref{eq:stability-cond}.
From this condition, the steady state points 
are found to fall into the three following categories:
(a)~the unstable points with $\Re\Lambda_i<0$; 
(b)~the stable points with $\Re\Lambda_i=\Lambda_i>0$
(these are the trapping points that remain stable even if $\gamma=0$); 
and 
(c)~the conditionally
stable (stabilizable) points with $\Re\Lambda_i\ne\Lambda_i>0$
(such points being unstable at $\gamma=0$ can be stabilized
provided the damping constant is sufficiently large).

Figures~\ref{fig:intensity-00}--\ref{fig:zeq-10}
present the results for
incident non-vortex LG beams
with vanishing azimuthal number, $m_{\ind{LG}}=0$,
and the focusing parameter $f=0.3$
($2\pi f=\lambda/w_0$).
The Gaussian (LG$_{00}$) and non-Gaussian (LG$_{10}$)
beams (the intensity distributions are shown
in Figs.~\ref{fig:intensity-00} and~\ref{fig:intensity-10},
respectively)
are both characterized by the diagonal
stiffness matrix~\eqref{eq:L-m0}
and stability of the equilibria is thus independent
of the ambient medium.
The longitudinal eigenvalue $\Lambda_z$
given by Eq.~\eqref{eq:lambda-z}
controls the axial stability of the equilibrium points
and all our results for the eigenvalues and the
location of equilibria are limited to the case of 
axially stable points with $\Lambda_z>0$.

Referring to Figs.~\ref{fig:eigval-00} and~\ref{fig:zeq-00},
this is the axial stability that determines
stability of the trapping points
depending on the size parameter $D_p/\lambda$
of the particle illuminated with the Gaussian beam.
By contrast, the results for the non-Gaussian LG$_{10}$ beam
shown in Figs.~\ref{fig:eigval-10} and~\ref{fig:zeq-10}
indicate that all the points are axially stable 
and their stability is governed by the size dependence
of the transverse eigenvalue $\Lambda_x$.

The
principal characteristic feature of 
the conservative radiation-force-induced dynamics
illustrated by the non-vortex LG beams
is that the stiffness matrix is symmetric
and its eigenvalues are real-valued.
Therefore,
such dynamics is characterized by the absence
of  conditionally stable points with $\Im\Lambda_i\ne 0$. 

We have found that, for purely azimuthal LG beams
with the vanishing radial number, $n_{\ind{LG}}=0$,
and the nonzero azimuthal mode number $m_{\ind{LG}}\in\{1,2\}$, the latter is no longer the case.
Such beams 
(the intensity distributions for the LG$_{01}$ and LG$_{02}$ beams
are shown
in Figs.~\ref{fig:intensity-01} and~\ref{fig:intensity-02},
respectively)
represent the case of optical vortex LG beams
carrying a phase singularity and exhibiting a helical phase front.

Equation~\eqref{eq:lambda-pm}
gives the transverse eigenvalues of
the stiffness matrix~\eqref{eq:L-on-axis}
for the optical vortex beams.
The eigenvalues computed as a function
of the size parameter for the LG$_{01}$ and LG$_{02}$ beams
are plotted in Figs.~\ref{fig:eigval-01} and~\ref{fig:eigval-02},
respectively.
These figures clearly indicate 
the intervals of the size parameter
where the equilibrium points are
conditionally stable with $\Im\Lambda_{\pm}\ne 0$
and $\Re\Lambda_{+}=\Re\Lambda_{-}>0$.
In both cases, at small values of the size parameter,
the transverse eigenvalues play the role of 
the destabilizing factor.
For the LG$_{01}$ beam,
similar to the Gaussian beam, 
stability of the equilibria
outside the region of small scatterers
is controlled by the longitudinal eigenvalue $\Lambda_z$.
When $m_{\ind{LG}}=2$,
the zero-force points are axially stable
and, similar to the case of the LG$_{10}$ beam, 
stability is determined by the transverse eigenvalues,
$\Lambda_{+}$ and $\Lambda_{-}$.

In figures showing 
the curves for the eigenvalues
and the equilibrium coordinate $z_{\ind{eq}}$,
we have used differently shaped symbols
to mark
the values
of the size parameter $D_p/\lambda$
corresponding to local maxima of 
the enhancement factors 
$A_j=|a_j^{(p)}|^2$
and $B_j=|b_j^{(p)}|^2$,
where $a_j^{(p)}$
and $b_j^{(p)}$
are the internal field coefficients.
For non-Gaussian LG beams,
the endpoints of the instability and conditional stability
intervals are found to be close to certain Mie resonance points.
Similar remark applies to 
the minima characterizing
oscillating behavior of
the size dependence of $z_{\ind{eq}}$
in the large size region. 
The resonances in
the Mie coefficients and the related
interference effects
are thus found to play the role of 
the factor
changing the trapping properties of the particles.
Similarly, the results of Ref.~\cite{Stilgoe:optexp:2008}
show that
the Mie resonances have a profound
effect on the trapping characteristics
of high refractive index particles where 
the interference effects are expected to be
strongest.


In conclusion, we note that
our symmetry considerations
tacitly assume that 
the incident beam is solely responsible for
breaking the spherical symmetry of the optically isotropic dielectric scatterer.
The symmetry can additionally be reduced by 
the optical anisotropy~\cite{Kiselev:pre:2002,Kiselev:mclc:2002}
that may thus significantly affect the regimes of the radiation-force-induced
dynamics of the particle.
Despite some recent results on the radiation force exerted 
on uniaxially anisotropic spheres~\cite{Li:optexp:2012,Qu:jqsrt:2015},
the optical anisotropy related effects are still far from
being well understood.

\begin{acknowledgments}
ADK acknowledges partial financial support
from the Government of the Russian Federation (Grant No.
074-U01), from the Ministry of Education and Science of
the Russian Federation (Grant No. GOSZADANIE 2014/190,
Project No. 14.Z50.31.0031, and ZADANIE Grant No.
1.754.2014/K), through a grant from the Russian Foundation
for Basic Research, and through a grant from the President of
Russia (Grant No. MK-2736.2015.2).
\end{acknowledgments}



%

\appendix

\section{Gradient terms in far-field expression for optical force}
\label{sec:derivation}

In this Appendix we consider
the case of non-absorbing
scatterer and show
how to rearrange the far-field expression
for the optical force~\eqref{eq:F_far-field}
so as to separate out 
the gradient part of the force.
For this purpose,
we begin with the far-field distribution
of the scattered wave~\eqref{eq:E_out_sca_expan} 
rewritten in the  following form: 
\begin{align}
&
  \label{eq:E_out_sca_in}
 \vc{E}_{\ind{out}}^{\ind{(sca)}}(\uvc{r},\vc{r}_p)
\equiv \vc{E}_{\ind{out}}^{\ind{(sca)}}
=
\sum_{jm}\sum_{\alpha\in\{e,m\}}
s_{jm}^{(\alpha)}(\vc{r}_p)\vc{Y}_{jm}^{(\alpha)}(\uvc{r})
=
2 \sum_{jm}\sum_{\alpha\in\{e,m\}}
T_{j}^{(\alpha)}
w_{jm}^{(\alpha)}(\vc{r}_p)\vc{Y}_{jm}^{(\alpha)}(\uvc{r})
\notag
\\
&
=
2 \avr{\mathcal{T}(\uvc{r},\uvc{r}')\vc{E}_{\ind{out}}^{\ind{(inc)}}(\uvc{r}',\vc{r}_p)}_{\uvc{r}'}
\equiv 2 \mathcal{T} \vc{E}_{\ind{out}}^{\ind{(inc)}},
\\
&
  \label{eq:T_oprt}
\mathcal{T}(\uvc{r},\uvc{r}')=
\sum_{jm}\sum_{\alpha\in\{e,m\}}
T_{j}^{(\alpha)}
\vc{Y}_{jm}^{(\alpha)}(\uvc{r})\otimes
\cnj{[\vc{Y}_{jm}^{(\alpha)}(\uvc{r}')]},
\end{align}
where $\mathcal{T}(\uvc{r},\uvc{r}')$ is the kernel of
the $T$-matrix operator $\mathcal{T}$; 
$T_j^{(m)}=T_j^{11}$ and $T_j^{(e)}=T_j^{22}$
are the Mie coefficients given by 
Eqs.~\eqref{eq:mie-alp-sca} and~\eqref{eq:mie-bet-sca},
respectively.
For non-absorbing particles,
the energy absorption rate~\eqref{eq:S_tot_flux}
vanishes and the $T$-matrix operator 
satisfies the unitarity relation:
\begin{align}
  \label{eq:T-unit-oprt}
  2 \hcnj{\mathcal{T}} \mathcal{T}+
\hcnj{\mathcal{T}}+\mathcal{T}=0.
\end{align}
The optical force then can be recast into the operator form:
\begin{align}
&
  \label{eq:F-opt-1}
  \vc{F}=-\frac{\epsilon}{4\pi k^2}
  \avr{\sca{\cnj{[\vc{E}_{\ind{out}}^{\ind{(inc)}}]}}{\mathcal{F}\vc{E}_{\ind{out}}^{\ind{(inc)}}}}_{\uvc{r}},
\\
&
\label{eq:F-opt-2}
\mathcal{F}=2 \hcnj{\mathcal{T}}\uvc{r}\mathcal{T}+
\uvc{r}\mathcal{T}+\hcnj{\mathcal{T}}\uvc{r}
\notag
\\
&
=
[\uvc{r},\mathcal{T}]+2 \hcnj{\mathcal{T}}\uvc{r}\mathcal{T}-
2 \hcnj{\mathcal{T}}\mathcal{T}\uvc{r}
=
[\hcnj{\mathcal{T}},\uvc{r}]+2 \hcnj{\mathcal{T}}\uvc{r}\mathcal{T}-
2 \uvc{r}\hcnj{\mathcal{T}}\mathcal{T},
\end{align}
where 
we have used the unitarity relation~\eqref{eq:T-unit-oprt} to 
transform the expression for the operator $\mathcal{F}$
and
$[\mathcal{A},\mathcal{B}]=\mathcal{A}\mathcal{B}-\mathcal{B}\mathcal{A}$
stands for the commutator of operators $\mathcal{A}$ and $\mathcal{B}$. 
From Eq.~\eqref{eq:shift_E_out}, it can readily be seen that
multiplication of the far-field vector amplitude 
$\vc{E}_{\ind{out}}^{\ind{(inc)}}(\uvc{r},\vc{r}_p)$ 
by the unit vector $\uvc{r}$
can be replaced with the following differential (gradient)
operation:
\begin{align}
  \label{eq:nbl-0}
  \uvc{r}\vc{E}_{\ind{out}}^{\ind{(inc)}}=\tilde{\bnbl}_p\vc{E}_{\ind{out}}^{\ind{(inc)}},
\quad \tilde{\bnbl}_p=-ik^{-1}\bnbl_p=-ik^{-1}
\left(\frac{\partial}{\partial x_p },\frac{\partial}{\partial y_p },\frac{\partial}{\partial z_p }\right).
\end{align}
Our next step is to derive the relations
\begin{align}
&
  \label{eq:nbl-1}
  -\avr{\sca{\cnj{[\vc{E}_{\ind{out}}^{\ind{(inc)}}]}}{[\uvc{r},\mathcal{T}]\vc{E}_{\ind{out}}^{\ind{(inc)}}}}_{\uvc{r}}
=\tilde{\bnbl}_p\avr{\sca{\cnj{[\vc{E}_{\ind{out}}^{\ind{(inc)}}]}}{\mathcal{T}\vc{E}_{\ind{out}}^{\ind{(inc)}}}}_{\uvc{r}}=
\tilde{\bnbl}_p\avr{\sca{\cnj{[\vc{E}_{\ind{out}}^{\ind{(inc)}}]}}{\vc{E}_{\ind{out}}^{\ind{(sca)}}}}_{\uvc{r}},
\\
&
  \label{eq:nbl-2}
  \avr{\sca{\cnj{[\vc{E}_{\ind{out}}^{\ind{(inc)}}]}}{\hcnj{\mathcal{T}}\mathcal{T}\uvc{r}\vc{E}_{\ind{out}}^{\ind{(inc)}}}}_{\uvc{r}}=
\avr{\sca{\cnj{[\vc{E}_{\ind{out}}^{\ind{(sca)}}]}}{\tilde{\bnbl}_p \vc{E}_{\ind{out}}^{\ind{(sca)}}}}_{\uvc{r}}
\end{align}
that immediately follow from Eq.~\eqref{eq:nbl-0}
since the $T$-matrix operator and the Mie coefficients are
both independent of the displacement vector $\vc{r}_p$.

Relations~\eqref{eq:nbl-1}--\eqref{eq:nbl-2}
and equation~\eqref{eq:F-opt-2}
can now be substituted into formula~\eqref{eq:F-opt-1}
to obtain the result in the final form:
\begin{align}
&  
\label{eq:F-grad}
  \vc{F}=\frac{\epsilon}{4\pi k^2}
\Bigl\{
- 2 \avr{\uvc{r}
  \sca{\cnj{[\vc{E}_{\ind{out}}^{\ind{(sca)}}]}}{\vc{E}_{\ind{out}}^{\ind{(sca)}}}}_{\uvc{r}}
\notag
\\
&
+
k^{-1}\Im
\Bigl[
{\bnbl}_p\avr{\sca{\cnj{[\vc{E}_{\ind{out}}^{\ind{(inc)}}]}}{\vc{E}_{\ind{out}}^{\ind{(sca)}}}}_{\uvc{r}}+
2 \avr{\sca{\cnj{[\vc{E}_{\ind{out}}^{\ind{(sca)}}]}}{{\bnbl}_p
    \vc{E}_{\ind{out}}^{\ind{(sca)}}}}_{\uvc{r}}
\Bigr]
\Bigr\},
\end{align}
where the last two terms on the right hand side of
Eq.~\eqref{eq:F-grad} represent 
a derivative dependent (gradient) contribution to 
the radiation force.
It should be emphasized that
the last term being generally non-conservative
will contribute to the asymmetry of the stiffness matrix.

\end{document}